\documentclass[prd,twocolumn,aps,preprintnumbers,letterpaper]{revtex4-1}
\usepackage{amsmath}
\usepackage{tipa}
\usepackage{amssymb}
\usepackage{graphicx}
\usepackage{xfrac}
\usepackage{bm}
\usepackage{enumerate}
\usepackage{color}
\usepackage[colorlinks = true,
            linkcolor = blue,
            urlcolor  = blue,
            citecolor = blue,
            anchorcolor = blue]{hyperref}
\usepackage{subfigure}
\usepackage{braket}
\usepackage{rotating}
\usepackage{capt-of}
\usepackage{chngcntr}

\newcommand{\dd}{\mathrm{d}}
\newcommand{\eq}[1]{\begin{equation}#1\end{equation}}
\newcommand{\al}[1]{\begin{align}#1\end{align}}
\newcommand{\nn}{\nonumber}
\newcommand{\Tr}{\mbox{Tr}\;}

\def\hri#1#2{\href{http://arxiv.org/abs/#1}{[arXiv:#1]}}
\def\hre#1#2{\href{http://arxiv.org/abs/#1/#2}{[arXiv:#1/#2]}}

\begin{document}

\title{Effective Model of QCD Magnetic Monopoles From Numerical Study\\ of One- and Two-Component Coulomb Quantum Bose Gases}
\author{Adith Ramamurti}
\email[]{adith.ramamurti@stonybrook.edu}
\author{Edward Shuryak}
\email[]{edward.shuryak@stonybrook.edu}
\affiliation{Department of Physics and Astronomy, Stony Brook University,\\ Stony Brook, New York 11794-3800, USA}

\date{\today}

\begin{abstract}

Magnetic monopoles are suggested to play an important role in strongly coupled quark-gluon plasma (sQGP) near the deconfinement temperature. So far, their many-body treatment has only been done classically, with just binary scattering solved in quantum mechanics. In this paper we start {\em quantum} many-body studies of the monopole ensembles. Specifically,  we carry out numerical simulations of the path integral for one- and two-component Coulomb Bose systems.  We determine the relation between the critical temperature for the Bose-Einstein condensation phase transition $T_\text{c}$ and the Coulomb coupling strength using two methods, the classic finite-size scaling of the condensate and a lattice-tested method based on permutation cycles. For a one-component Coulomb Bose gas, we observe the same behavior of the critical temperature -- initially rising slightly then falling as interaction strength is increased -- as seen in the case of hard spheres; we also observe the same behavior for a two-component Coulomb Bose gas. We then calculate sets of radial correlation functions between the like and unlike charged particles. By matching those with  the correlation functions previously calculated on the lattice, we derive an effective quantum model of color magnetic monopoles in QCD.  From this matched model, we are able to extract the monopole contribution to QCD equation of state near $T_\text{c}$. 

\end{abstract}

\maketitle

\section{Introduction}

Dirac \cite{Dirac60} showed that magnetic monopoles can exist, provided the product of the electric and magnetic couplings is an integer multiple of $4 \pi$, which makes the \textit{Dirac strings} invisible. A specific solution for the monopole has been found by `t Hooft \cite{NUPHA.B79.276} and Polyakov \cite{Polyakov:1974ek} for a gauge theory with an adjoint colored scalar, known as the Georgi-Glashow model. 

In the 1970s, Nambu \cite{Nambu:1974zg}, `t Hooft \cite{NUPHA.B190.455}, and Mandelstam \cite{Mandelstam:1974pi} proposed the dual superconductivity model of the QCD vacuum, suggesting that confinement is due to the Bose-Einstein condensation (BEC) of magnetic monopoles.

With the advent of numerical simulations on the lattice, this scenario has been tested in multiple ways and essentially confirmed. For example, it was found that, at $T<T_\text{c}$, magnetic monopoles rotate around the electric flux tubes, producing a supercurrent. These monopoles have the same properties as electric charges rotating around magnetic flux tubes in an ordinary superconductor, hence the name ``dual." Specific properties of the monopoles -- the correlations, densities, and condensates -- have also been evaluated on the lattice, c.f. \cite{DAlessandro:2007lae, D'Alessandro:2010xg, Bonati:2013bga, Chernodub:2009hc}. 

The ``magnetic scenario" for finite-temperature QCD \cite{Liao:2006ry} suggested that interactions of the ``electric" objects -- quarks and gluons -- with magnetic monopoles give rise to the unusual transport properties of the quark-gluon plasma (QGP).  This idea was first developed via classical molecular dynamics simulations of a ``dual plasma," containing electrically and magnetically charged particles. These simulations, along with the lattice studies mentioned above, showed that the magnetic coupling runs as Dirac predicted: as temperature is increased, the magnetic coupling {\em grows} inversely to the electric coupling \cite{PhysRevLett.101.162302}.

Monopoles have subsequently been included in calculations of transport properties via quantum-mechanical  binary scattering amplitudes -- see, for example, \cite{Ratti:2008jz} for gluon-monopole scattering. Another application of scattering of charges on monopoles is in models of jet quenching \cite{Xu:2014tda,Xu:2015bbz}.

The purpose of this work is to elevate the classical dual Coulomb plasma picture to an effective quantum many-body theory of monopole ensembles. The need for a quantum model is clear: without one, it would not be possible to study the Bose-Einstein condensation transition.
 
Lattice simulation of gauge theories include magnetic monopoles as certain solitons made of glue. These simulations are based on first principles, namely the QCD Lagrangian. However, they also include many more degrees of freedom -- such as quark and gluon quasiparticles at $T>T_\text{c}$ -- and are therefore very expensive. Our aim is to create an effective model of the monopoles and quantify their contributions to various observables. In doing so, we realize that one can only separate the monopoles from other degrees of freedom to a certain degree.

To simulate these quantum Coulomb Bose systems, we will use Path-Integral Monte Carlo (PIMC). This method has been widely used since the 1980s;  for extensive detail and an overview of its early successes, see \cite{RevModPhys.67.279}. The analysis methods we use for our simulations will be briefly discussed in Sec. \ref{methods}. 

We will first investigate how the Coulomb interaction between the  magnetic quasiparticles affects the critical temperature $T_\text{c}$ of their BEC phase transition, through numerical simulations of one- and two-component (plus and minus charged) Bose gases at different coupling strengths. The results of our simulations are in Sec. \ref{numsimres}. In Sec. \ref{eff}, we will map the results of our simulations for the two-component Bose gas to those found on the lattice in order to find the parameters for our model that yield the same effective behavior of the magnetic monopoles, and make estimates of the monopole contribution to QCD thermodynamics.

\section{Numerical Simulation Methodology}\label{methods}

\subsection{Path Integral Monte Carlo}\label{PIMC}

The path-integral formulation of quantum mechanics was developed by Feynman \cite{Feynman:1948ur}, who also extended this formalism to describe statistical mechanics, using periodic path integrals in Euclidean time. For many years this formalism has been used for perturbation theory, the derivation of Feynman diagrams at zero- and finite-temperature, and analytic semiclassical methods. 

One of Feynman's early applications of the path integral formalism was to the BEC phenomenon, which he connected to to the appearance of Bose-clusters of particles; as the temperature drops from above $T_\text{c}$ to below, the suppression of these clusters disappears. He qualitatively explained why an interacting quantum system may have a lower Bose-Einstein condensation critical temperature than in the case of free particles, famously applying this method to study liquid $^4He$ at near-zero temperatures, \cite{PhysRev.91.1291, Feynman:1953zz}.

Numerical evaluation of path integrals became feasible in late 1970s. Particularly, after pioneering work by Creutz  on confinement in lattice gauge theory \cite{PhysRevLett.43.553}, Creutz and Freedman numerically computed path integrals for quantum-mechanical motion in quartic potential \cite{Creutz:1980gp}. Some further examples of simulations with a few particles, such as two electrons in the the $He$ atom and the four nucleons in a $^4He$ nucleus, were done by one of us \cite{Shuryak:1984xr}. The basis for all of these simulations is the Metropolis algorithm, \cite{Metropolis:1953am}. Starting in the early 1980s, supercomputers allowed numerically simulate quantum many-body systems, such as liquid $^4He$. 

For self-consistency of the paper, we present a brief summary of the relevant general formulae and methods used in Appendix \ref{app_intro}. Here, we will briefly describe two methods of finding the critical condensation temperature to be used below. These are more feasible than the brute force approach, based on a calculation of the free energy of the ensemble with subsequent determination of specific heat and its peak. 

 The first is the method outlined by Pollock and Ceperley \cite{PhysRevB.36.8343}, and Pollock and Runge \cite{PhysRevB.46.3535}, based on supercurrent and its finite-size scaling. The second method, developed by Cristoforetti and Shuryak \cite{Cristoforetti:2009tx}, uses the permutation-cycle statistics of the system to find $T_\text{c}$. The latter method has not been used in analysis of PIMC simulations, so this work also seeks to test the accuracy of this method. It has so far been used in \cite{D'Alessandro:2010xg} and \cite{Bonati:2013bga} for lattice monopoles, confirming that deconfinement $T_\text{c}$ is indeed the BEC transition of the monopoles.

\subsubsection{$T_\text{c}$ from finite-size scaling of the superfluid fraction}

Following the discussion in \cite{PhysRevB.36.8343, PhysRevB.46.3535}, we can identify the winding number of the system with the superfluid fraction, and then using finite-size scaling, determine $T_\text{c}$. In experimental settings, the normal and superfluid components of a system are determined from boundary behavior. If we introduce a velocity $v$ to the boundaries of our system, we have a new density matrix
\eq{
\rho_v = \exp\{\beta H_v\}\,,
}
with 
\eq{
H_v = \sum_j \frac{({\bf p}_j-m{\bf v})^2}{2m} + V \,.
}

The normal component of the fluid is the portion that responds to this boundary motion, so we can write for the total momentum,
\eq{
\frac{\rho_N}{\rho} Nm{\bf v} = \braket{\textbf{P}}_v \,.
}
We have for the free energy of this system,
\eq{
\exp\{- \beta F_v\}  = \Tr{\rho_v}\,,
}
so we can write 
\eq{
\frac{\rho_N}{\rho} Nm{\bf v}  = -\frac{\partial F_v}{\partial \textbf{v} } + N m \textbf{v}\,,
}
or, equivalently, 
\eq{\label{df}
\frac{\rho_s}{\rho} = \frac{\partial(F_v/N)}{\partial(\frac{1}{2}mv^2)} \rightarrow \frac{\Delta F_v}{N} =\frac{1}{2}mv^2 \frac{\rho_s}{\rho}  + ...
}

In the path integral formalism, the density matrix with a velocity obeys the Bloch equation with moving walls, with periodic boundary conditions such that it is identical with a translation by a lattice vector. We can define a transformed density matrix, $\rho'$, by
\eq{
\rho_v(R,R';\beta) = \exp\left\{i \frac{m}{\hbar}\textbf{v}\cdot\sum_j(\textbf{r}_j - \textbf{r}'_j)\right\}\rho'(R,R';\beta)\,.
}

This new density matrix obeys the Bloch equation in the case of \textit{stationary walls}, but obtains a factor of $\exp\left\{i \frac{m}{\hbar}\textbf{v}\cdot\textbf{L}\right\}$ in periodic translations. Keeping track of the number of times the periodic boundary conditions are applied can be done with the definition of a \textit{winding number}, which counts the number of times a particle winds around the spatial directions of periodic box before returning to its ``original" location.

The free energy change induced from a velocity $v$ can be written as
\eq{
\exp\left\{\beta \Delta F_v\right\} = \exp\left\{i \frac{m}{\hbar}\textbf{v}\cdot\textbf{W}L\right\}\,,
}
\eq{
\beta \Delta F_v = \frac{m^2 v^2}{2 \hbar^2} \frac{\braket{W^2}L^2}{3} + ...
}
We can thus identify, with use of Eq. (\ref{df}),
\eq{
\frac{\rho_s}{\rho} = \frac{m}{\hbar} \frac{\braket{W^2}L^2}{3\beta N} \,.
}

From assumptions of finite-size scaling, we have that, near $T_\text{c}$,
\eq{
\frac{\rho_s}{\rho} (T, L) = L^{-1} Q(L^{-1/\nu}t)\,,
}
with $t= (T-T_\text{c})/T_\text{c}$. As a result, the functions $L\rho_s/\rho (T, L)$, for different values of $L$, should all cross at $T_\text{c}$, barring some minor corrections, expanded upon in \cite{PhysRevB.46.3535}.

\subsubsection{$T_\text{c}$ from permutation-cycle statistics}

Following the example of \cite{D'Alessandro:2010xg}, the partition function of a \textit{non-interacting} ideal gas of bosons can be broken up into a product of contributions of $k$-cycles -- where, for example, the permutation $(1,2,3)\rightarrow(2,3,1)$ is considered a 3-cycle -- as
\eq{
Z = \frac{1}{N!}\sum_P\prod_k z_k^{n_k} \,,
}
where $n_k$ is the number of $k$-cycles present in the system. Feynman's idea was that the sum over Bose-cluster size $k$ of the density of $k$-cycles should diverge at $T_\text{c}$, which in turn implies that some critical action $S^*$ required to permute two particles should be reached. He justified this idea for an ideal gas, and in \cite{Cristoforetti:2009tx}, it was extended to interacting systems such as liquid $^4 He$.

Expanding these contributions, 
\al{
z_k(T) &= \int \dd y_1 .. \dd y_k \braket{y_2,y_3,...,y_k,y_1|e^{-\beta \hat{H}}| y_1,y_2,...,y_k} \nonumber \\
&=  \int \dd y_1 \braket{y_1| e^{-k \beta \hat{H}}|y_1} \equiv z_1(T/k) \,.
}
The partition function for a non-relativistic free particle in a box is well known, so we can get the full contribution
\eq{
z_k(T) = \frac{V}{\lambda_B^3 k^{3/2}} \,,
}
where $\lambda_B$ is the thermal de Broglie wavelength and $V$ is the volume of the box. Then the partition function is
\eq{
Z = \frac{1}{N!} \sum_P \sum_k\left(\frac{V}{\lambda_B^3 k^{3/2}}\right)^{n_k}\,.
}
This quantity is not easily computed for fixed particle number, but this problem is avoided if we go to the Grand Canonical ensemble, for which the partition function is
\eq{
\mathcal{Z} = \prod_k\left(\frac{V e^{\mu k/T}}{\lambda_B^3 k^{5/2}}\right)\,.
}
From this partition function, we can extract the density of $k$-cycles
\eq{
\rho_k(T) \equiv \frac{\braket{n_k}}{V} = \frac{e^{\mu k /T}}{\lambda_B^3 k^{5/2}} \,.
}

The total particle density is
\eq{
\frac{N}{V} = \sum_k k \rho_k(T) = \sum_k \frac{e^{\mu k /T}}{\lambda_B^3 k^{5/2}} \,,
}
which has an upper limit of $\mu=0$. This approach is fully valid for any non-interacting gas above $T_\text{c}$ (i.e. with $k$-cycles on the microscopic scale). Therefore, by measuring numerically the densities $\rho_k$ at various temperatures of a given system, we can fit a curve of the function above to find a temperature at which the quantity $\mu/T$ vanishes; this value will be the critical temperature, $T_\text{c}$, for Bose-Einstein condensation (BEC).

In this paper, as was studied in \cite{D'Alessandro:2010xg, Bonati:2013bga}, we are \textit{not} dealing with a non-interacting gas of particles, for which the approach above is exact. Nevertheless, we expect the densities of cycles to decrease exponentially with $k$
\eq{
\rho_k(T) = e^{-\hat{\mu} k} f(k) \,, \label{eq_fit}
}
where $\hat{\mu} = - \mu/T$ and $f(k)$ is some decreasing function of $k$ of the form $f(k) \sim 1/k^\alpha$. To find the critical temperature, we look for the temperature the $k$-cycles are no longer suppressed exponentially in $k$, i.e. $\hat{\mu} \rightarrow 0$.

\section{Numerical Simulation Results}\label{numsimres}

\subsection{Test Case: BEC Critical Temperature for $^4He$}

The first task was to reproduce well-quantified results using the permutation-cycle critical temperature analysis, in order to ensure the validity and applicability of this method. To do so, we simulated a box of 128 $^4He$ particles interacting via the empirical Aziz potential \cite{Aziz:1979}, and compared our results to the experimental results as well as previous computational results  \cite{RevModPhys.67.279}. The experimental result for the BEC critical temperature was found to be 2.17 K, while the 
calculations with the binary Aziz potential predict a critical temperature of 2.19 K.

In order to compute the critical temperature using the permutation-cycle method, we first determine the temperature dependence of the probability of finding a particle in a $k$-cycle $P_k(T)$. From these probabilities, we compute the permutation-cycle densities, $\rho_k(T)$,
\eq{
\rho_k(T) = \frac{N P_k(T)}{k V} \,,
}
where $N$ is the number of particles in the system, $V$ is the volume.

\begin{figure}[h!]
\begin{center}
\subfigure[]{%
\includegraphics[width=\columnwidth,angle=0]{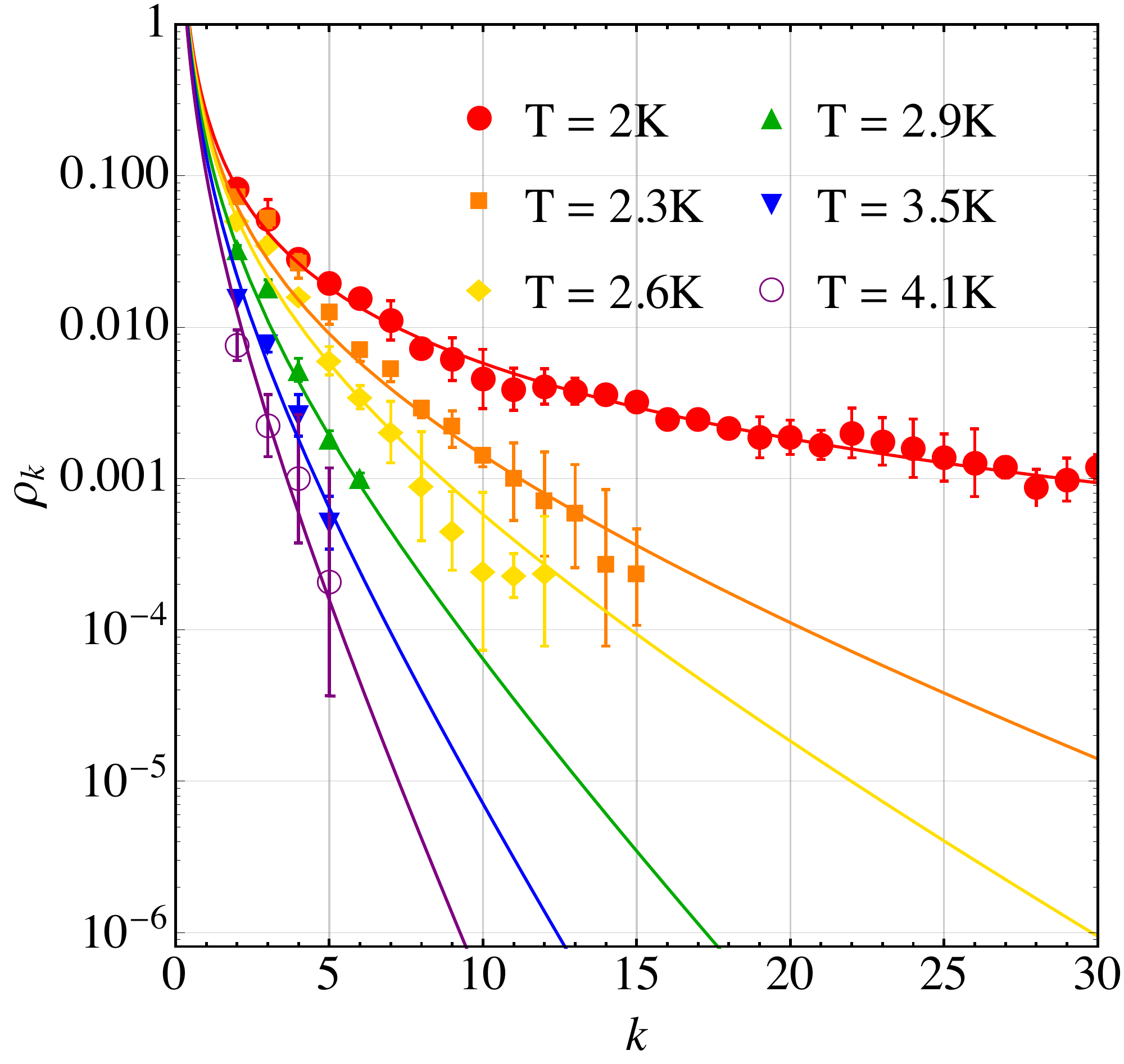}}\quad
\subfigure[]{%
\includegraphics[width=\columnwidth,angle=0]{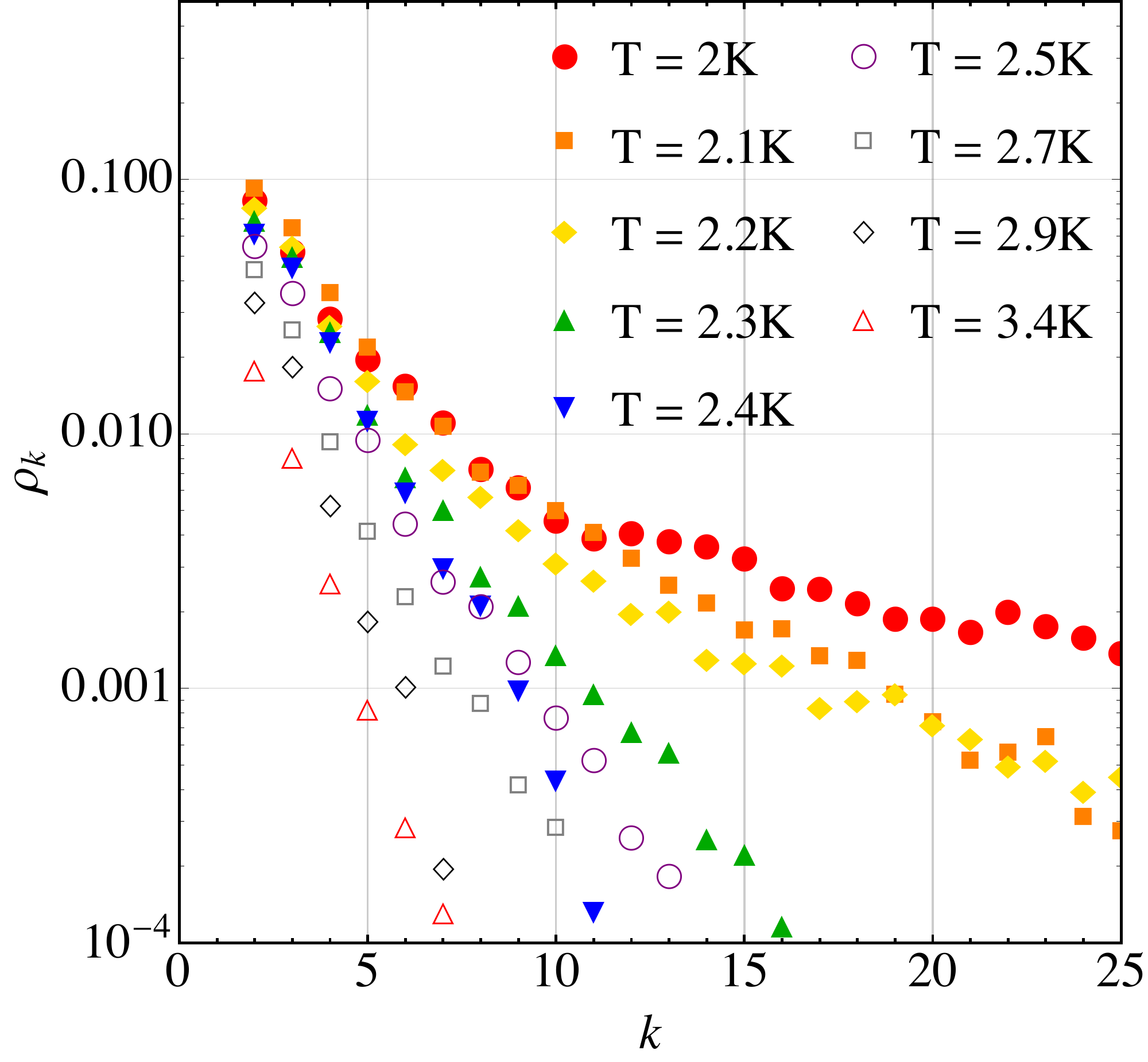}}\quad
\caption{The permutation cycle densities $\rho_k$ at various temperatures for a system of 128 $^4He$ particles. Fig. (a) shows a wider range of temperatures. The lines are rough fits to guide the eye. (b) focuses on temperatures near the critical value (neglecting error bars for clarity). Note the gap between the 2.2K and 2.3K  points.}
\label{fig_he4rhok}
\end{center}
\end{figure}

The permutation-cycle densities for various temperatures are shown in Fig. \ref{fig_he4rhok}. We can then fit these densities via Eq. (\ref{eq_fit}) to extract the suppression factor, $\hat{\mu}$. A few of these fitted curves are shown in Fig. \ref{fig_he4rhok}(a). At $T=2$ K -- below $T_\text{c}$ -- the exponential suppression is not present, and thus observe permutation cycles with $k>30$. This is a sign that at this temperature in an infinite system, there will be a cluster of infinitely many particles; the Bose condensate is present. In Fig. \ref{fig_he4rhok}(b), one can see explicitly how the exponential suppression appears at $T>2.2$ K, visually portrayed by the gap between the 2.2 K and 2.3 K lines. Above $T_\text{c}$, the suppression factor grows larger with temperature. 

\begin{figure}
\begin{center}
\includegraphics[width=\columnwidth,angle=0]{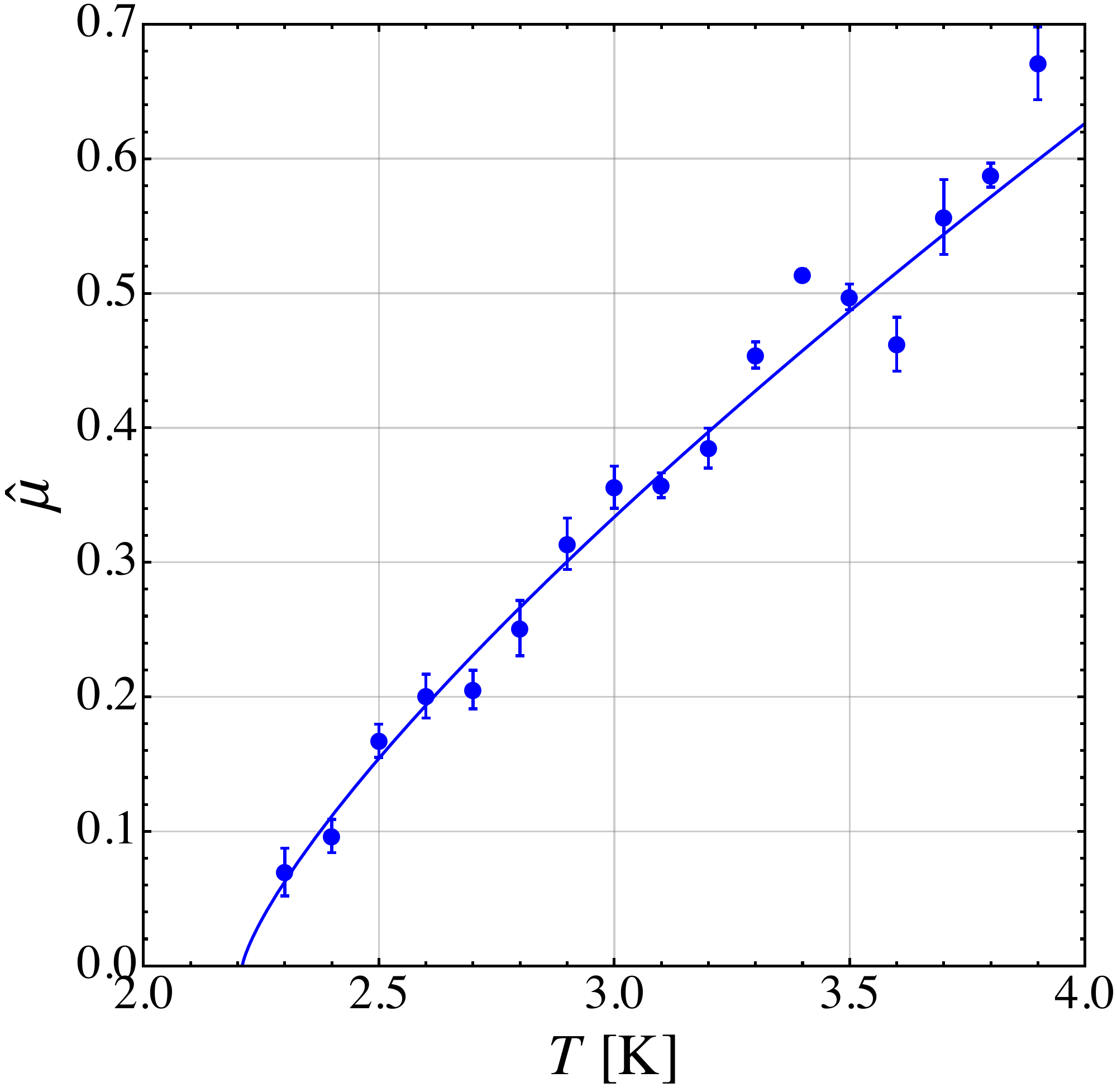}
\caption{The exponential suppression of $k$-cycles as a function of temperature for the $^4He$ system. The vanishing of the effective chemical potential $\hat \mu$ indicates the BEC critical temperature $T_\text{c}$.}
\label{fig_he4cond}
\end{center}
\end{figure}

Finally, after fitting all of the temperatures and finding $\hat{\mu}(T)$, we fit the $\hat{\mu}$ data with the functional form,
\eq{
\hat{\mu}(T) = A(T-T_\text{c})^\nu\,,
}
from which we find $T_\text{c}$. The results of our simulations are seen in Fig. \ref{fig_he4cond}. Using the permutation-cycle method, we find a critical temperature of 2.21$\pm$0.04 K. This result is within 2\% of reproducing the experimental critical temperature of the $^4He$ system and within 1\% of the critical temperature determined for the Aziz potential used in other numerical calculations.  We conclude that this method can indeed be used in path-integral Monte Carlo to accurately find the critical temperature of interacting Bose systems.

\subsection{BEC Critical Temperature for Coulomb Bose Gases}\label{restemp}

According to Einstein, the BEC of an ideal Bose gas happens at the critical temperature,
\eq{
T_\text{c}
 =    \left(\frac{2\pi\hbar^2}{m k_\text{B}}\right)\left( \frac{n}{\zeta\left(\frac{3}{2}\right)}\right)^\frac{2}{3}\,.
}
where $n$ is the density and $m$ is the particle mass. 

Extension of this relation to interacting Bose gases has an interesting history. There was much debate in the literature -- using Hartree-Fock, loop diagram, and renormalization group calculations, for example -- about even the sign of corrections to $T_\text{c}$; see \cite{PhysRevLett.79.3549} for discussion and references. 

Numerically, the dependence of BEC critical temperature on the strength of a hard sphere potential was studied by Gr\"{u}ter, et al.  \cite{PhysRevLett.79.3549}. It was found that at low densities the critical temperature is increased by a repulsive interaction, while at high densities the critical temperature is decreased, eventually recovering the well known $^4He$ result. This behavior at low densities can be explained with the calculation by Holzmann, et al. \cite{PhysRevLett.87.120403}.  

As a first step in making an effective model for a quantum two-component Coulomb Bose gas, we seek to find the dependence of $T_\text{c}$ on the Coulomb interaction strength; i.e., by varying $\alpha$ in
\eq{
V_\text{int}(r_{ij}) = \alpha \frac{q_i q_j}{r_{ij}} \,.
}
In our numerical study, the magnitude of charges, $q$; the masses, $m$; $\hbar$; Boltzmann's constant, $k_b$; and the density, $n$, are all scaled to 1. This leaves as variables only the temperature, $T$, entering via the period of the Euclidean time $\tau\in[0,1/T]$, the magnitude of the Coulomb coupling, $\alpha$, and the signs of the charges. In these units, the critical temperature for the ideal Bose gas is
\eq{
T_0 =    2\pi\left( \frac{1}{\zeta\left(\frac{3}{2}\right)}\right)^\frac{2}{3} = 3.3125\,.
}
and this value will be indicated by a horizontal dashed line in the plots to follow. More details of the numerical simulations explained hereafter, including system sizes and parameters, are given in Appendix \ref{app_details}.

\subsubsection{One-component Bose gas with varied Coulomb coupling parameter}

For the one-component Coulomb Bose gas, with compensating distributed charge commonly known as \textit{jellium}, we seek to investigate the dependence of $T_\text{c}$ on the strength of the Coulomb coupling parameter, and compare, qualitatively, it to the relation in the hard-sphere case.

\begin{figure}
\begin{center}
\includegraphics[width=\columnwidth,angle=0]{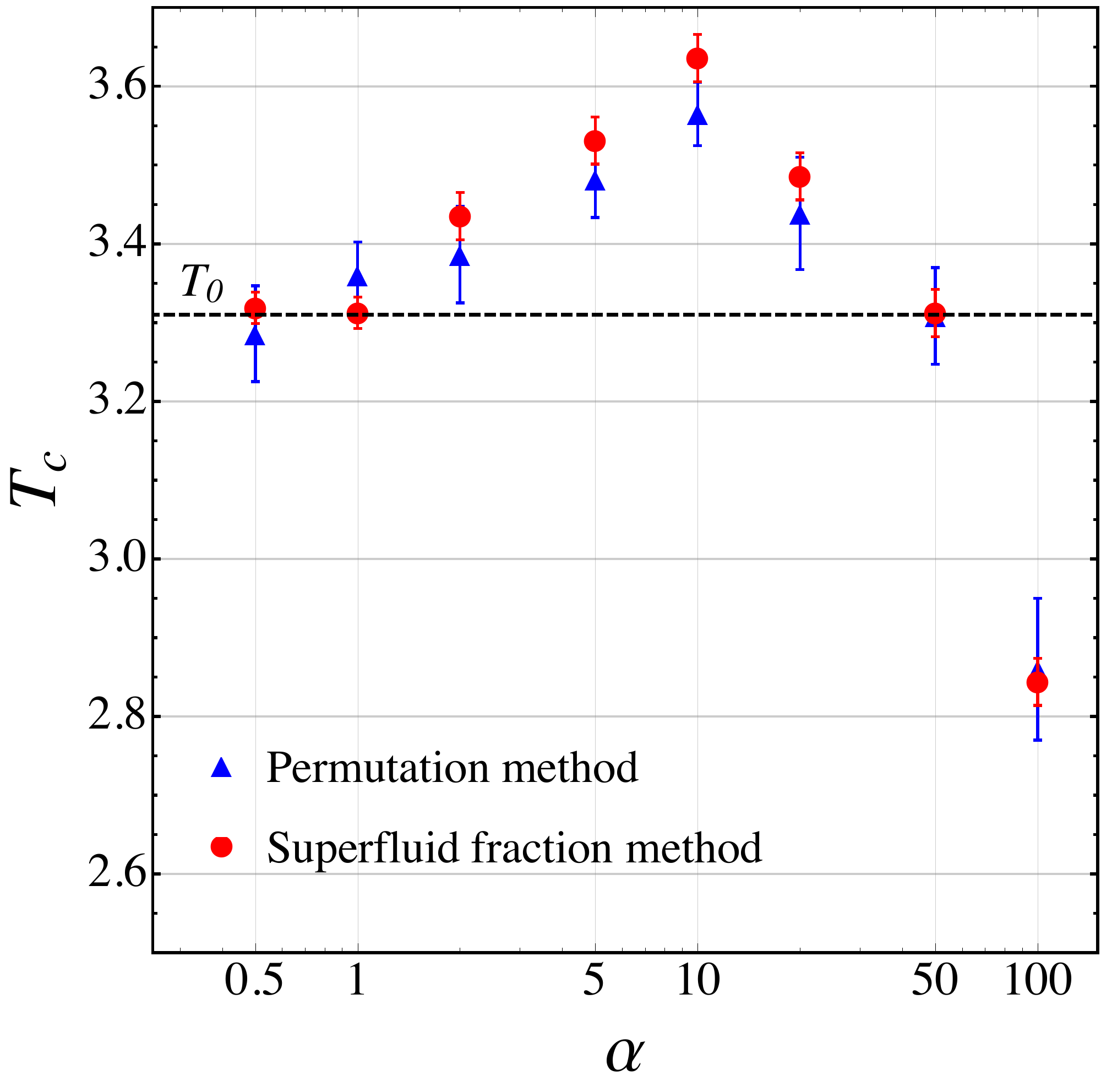}
\caption{The critical temperature for the BEC phase transition as a function of the coupling, $\alpha$. The red circles are the results of the finite-size scaling superfluid fraction calculation for systems of 8, 16, and 32 particles; and the blue triangles are the results of the permutation-cycle calculation for a system with 32 particles. The black dashed line denotes the Einstein ideal Bose gas critical temperature, $T_0$.}
\label{fig_onecomp}
\end{center}
\end{figure}

The results of our simulations are shown in Fig. \ref{fig_onecomp}. The first thing to note is that the two methods used produce results consistent within the statistical errors. Note further that we find the same behavior at small values of the coupling as in the case of low-density hard spheres \cite{PhysRevLett.79.3549}; the critical temperature for the BEC phase transition \textit{grows}. Yet if the coupling becomes large enough, $T_\text{c}$ rapidly drops below the critical temperature for an ideal Bose gas. Eventually, as the particles are ``too repulsive," the BEC phenomenon becomes impossible since it becomes essentially ``too costly" (in terms of the action, as compared to Feynman value) to permute them.

Let us also note that, while the permutation-cycle method agrees well with the older finite-size scaling method, the requirements for the system size to yield comparable results are different. The finite-size scaling method can give decent results even using two systems,  of only 8 and 16 particles, while the permutation cycle method required many runs of at least 32 particles. Therefore, at least in the case of long-range forces, which take a large amount of CPU time to compute, the finite-size scaling method may be more practical. If one, however, is looking at other quantities that require larger system sizes to begin with -- such as the superfluid fraction itself (and not just how it scales with system size) -- the permutation cycles method is an easy way of determining $T_\text{c}$ with data already gathered from those larger system simulations.

\begin{figure}
\begin{center}
\includegraphics[width=\columnwidth,angle=0]{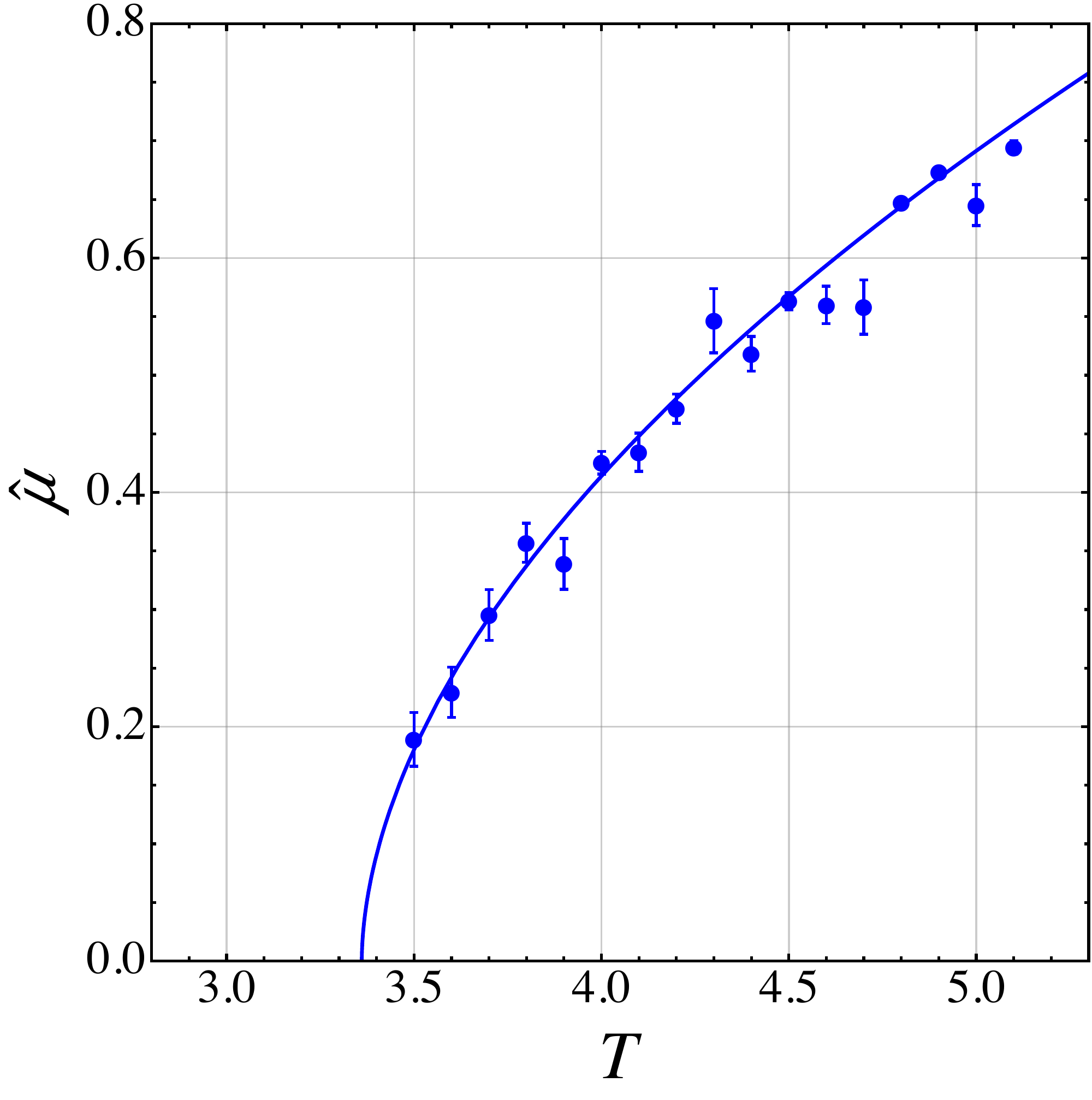}
\caption{The results of the permutation-cycle calculation for a system with 32 particles at $\alpha = 1$ along with a fitted curve. The vanishing of the effective chemical potential $\hat \mu$ indicates the critical temperature $T_\text{c}$. }
\label{fig_permtc1}
\end{center}
\end{figure}

\begin{figure}
\begin{center}
\includegraphics[width=\columnwidth,angle=0]{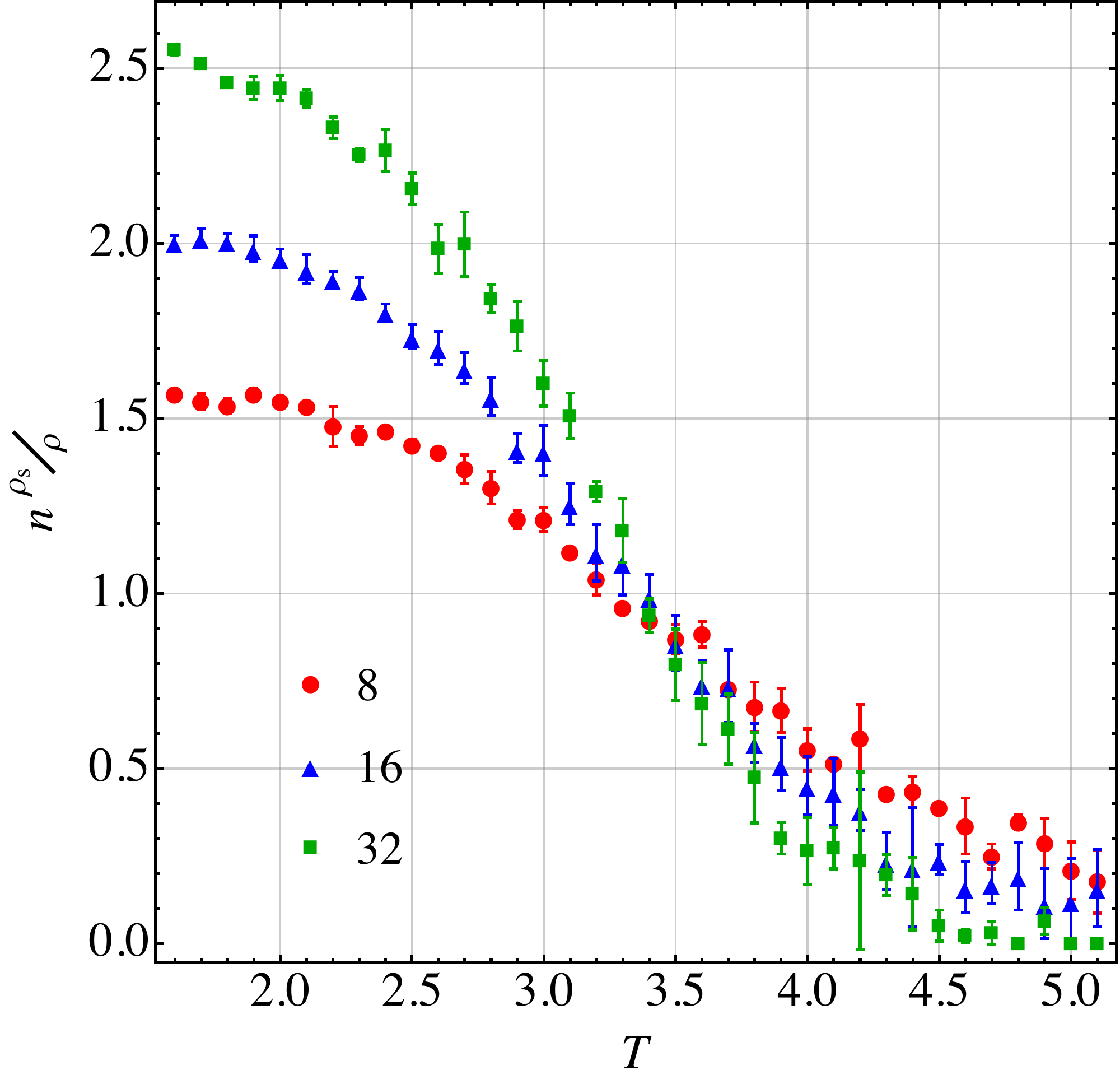}
\caption{The results of the finite-size scaling of the superfluid fraction calculation for 8 (red circles), 16 (blue triangles), and 32 (green squares) particles at $\alpha = 5$. The point where the data sets intersect is the critical temperature $T_\text{c}$.}
\label{fig_fsstc1}
\end{center}
\end{figure}

Examples of the analysis methods are shown in Figs. \ref{fig_permtc1} and \ref{fig_fsstc1}. In Fig. \ref{fig_permtc1}, we see the $\hat{\mu}$ data for the 32 particle system at $\alpha = 1$, which were obtained by fitting the permutation cycles as explained above. The $\hat{\mu}$ data is then fitted to find $T_\text{c}$, which is where the solid fitted curve intersects the $x$-axis. In Fig. \ref{fig_fsstc1}, we see the superfluid fraction data for three system sizes at $\alpha = 5$. The data is linearly fit around the intersection point, as described above. 

\subsubsection{Two-component Bose gas with varied Coulomb coupling }

\begin{figure}
\begin{center}
\subfigure[]{%
\includegraphics[width=0.47\textwidth]{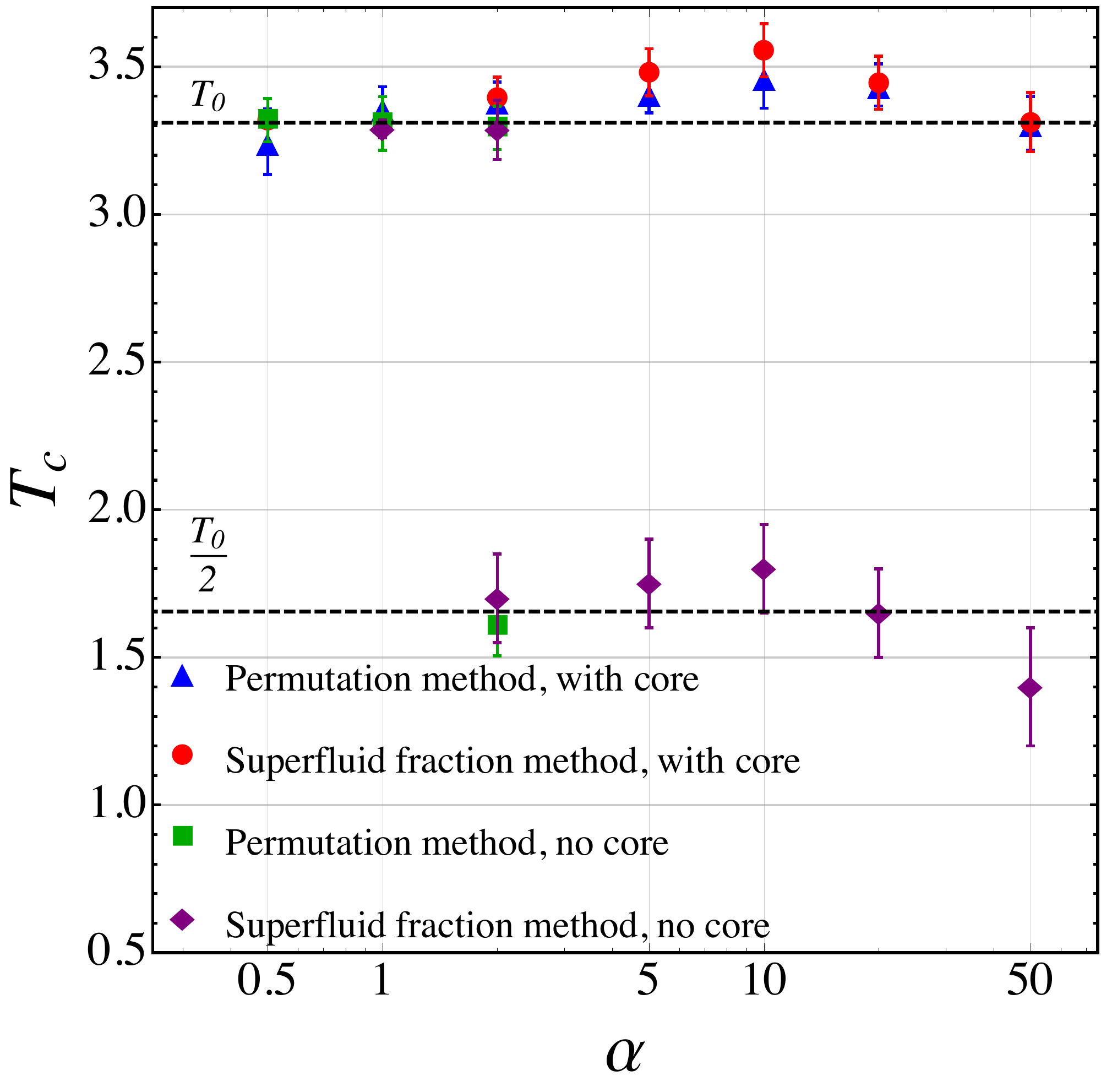}} \quad
\subfigure[]{%
\includegraphics[width=0.47\textwidth]{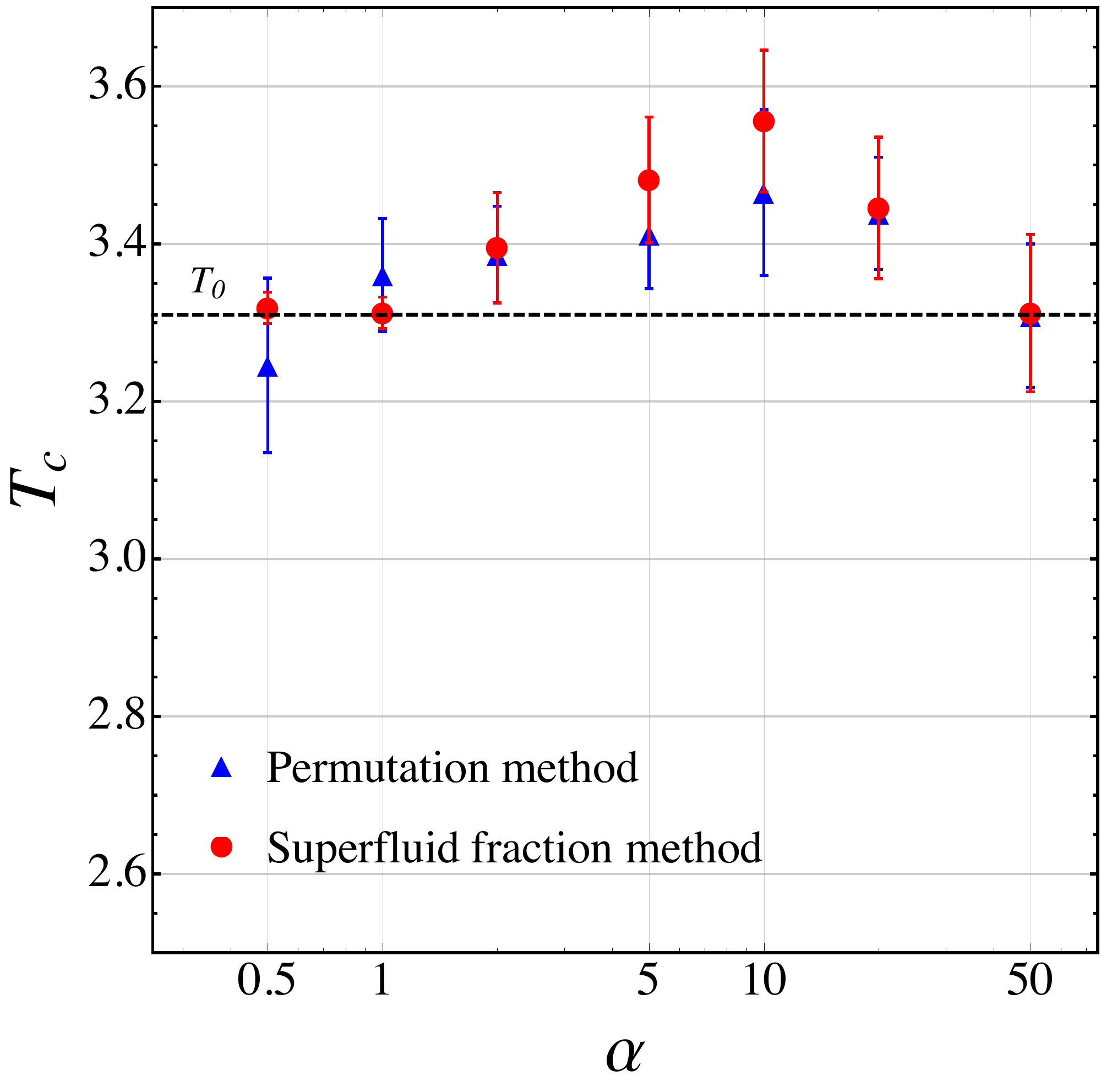}} \quad

\end{center}

\caption{The critical temperature for the BEC phase transition for the two-component Bose gas as a function of the coupling, $\alpha$. (a) The red circles and the blue triangles are the results of the finite-size scaling superfluid fraction calculation and the permutation-cycle calculation, respectively, for systems without core repulsion.. The purple diamonds and green squares are the results of the finite-size scaling superfluid fraction calculation and the permutation cycle calculation, respectively, for systems with core repulsion. The black dashed lines denote the Einstein ideal Bose gas critical temperature for a particle of mass $m=1$, $T_0$, and for a particle of mass $m=2$, $T_0/2$. (b) A zoom-in of the results in (a) for systems with core repulsion.}

\label{fig_twocomp}
\end{figure}

Following the  one-component Coulomb Bose gas, we carried out the PIMC simulations and analysis  for a neutral system of particles with two different charges, $+1$ and $-1$, with and without  repulsive core; the results are shown in Fig. \ref{fig_twocomp}. 

At very small couplings with and without core repulsion, the behavior of the critical temperature is very similar to that of the one-component case; the particles are only slightly interacting and therefore have a critical temperature very close to that of an ideal gas. When the coupling increases, however, the point-like (without core) two-component gas has a split in critical temperature, which acts like the one-component case for some particles, and drops much more quickly than that of the one-component case for others to approximately $T_0/2$, which is the ideal gas condensation temperature for particles with mass $m=2$, i.e., a neutral molecule comprised of particles with two opposite charges. 

The finite-size (with core) two-component gas, seen more clearly in Fig. \ref{fig_twocomp}(b), does not form these molecules and thus has the behavior of the one-component gas. It is important to note that even with a second component added, the finite-size particles have the {\em same} critical temperature behavior as a one-component gas: a 5-10\% deviation upward before dropping at very strong couplings.

 At extremely high couplings, the critical temperature drops for both cases. We do not have the accuracy in our data to confidently state whether the molecular phase has the same rising $T_\text{c}$ at lower couplings that is seen in the one-component case.

The data seen in Fig. \ref{fig_twocomp} for the case of point-like particles is from both the permutation cycle method and the finite-size scaling method for small couplings, but only finite-size scaling method for larger couplings, as there was too much statistical noise in the permutation statistics to obtain a good fit for a critical temperature. This may be remedied by a larger sample size or a larger system size, as the particles at higher coupling tend to become ``stuck" if they enter the molecule phase. We do not have this problem when the particles are given a repulsive core.

\subsection{Spatial Correlations}\label{rescorr}

One of the key observables we can use when comparing our results to the lattice monopole results is the spatial correlations of particles. In our simulations, we keep track of the Euclidean-time paths of individual particles, allowing us to observe their distributions in relative distance $r$ as a function of coupling strength and temperature. We define the pair correlation, $g(r)$, to satisfy
\eq{
n(r)=4\pi \rho \int _{0}^{r}g(r')r'^{2}dr'\,,
}
where $n(r)$ is the number of particles found between 0 and $r$, and $\rho = N/V$ is the overall density of particles in the volume; $g(r)$ is, by definition, normalized to distribution of an ideal gas. In this work, for cases in which we have two components, we denote the same charge correlation with $g_{++}(r)$ and the opposite charge correlation with $g_{+-}(r)$.

\begin{figure*}
    \centering
    \includegraphics[width=\hsize]{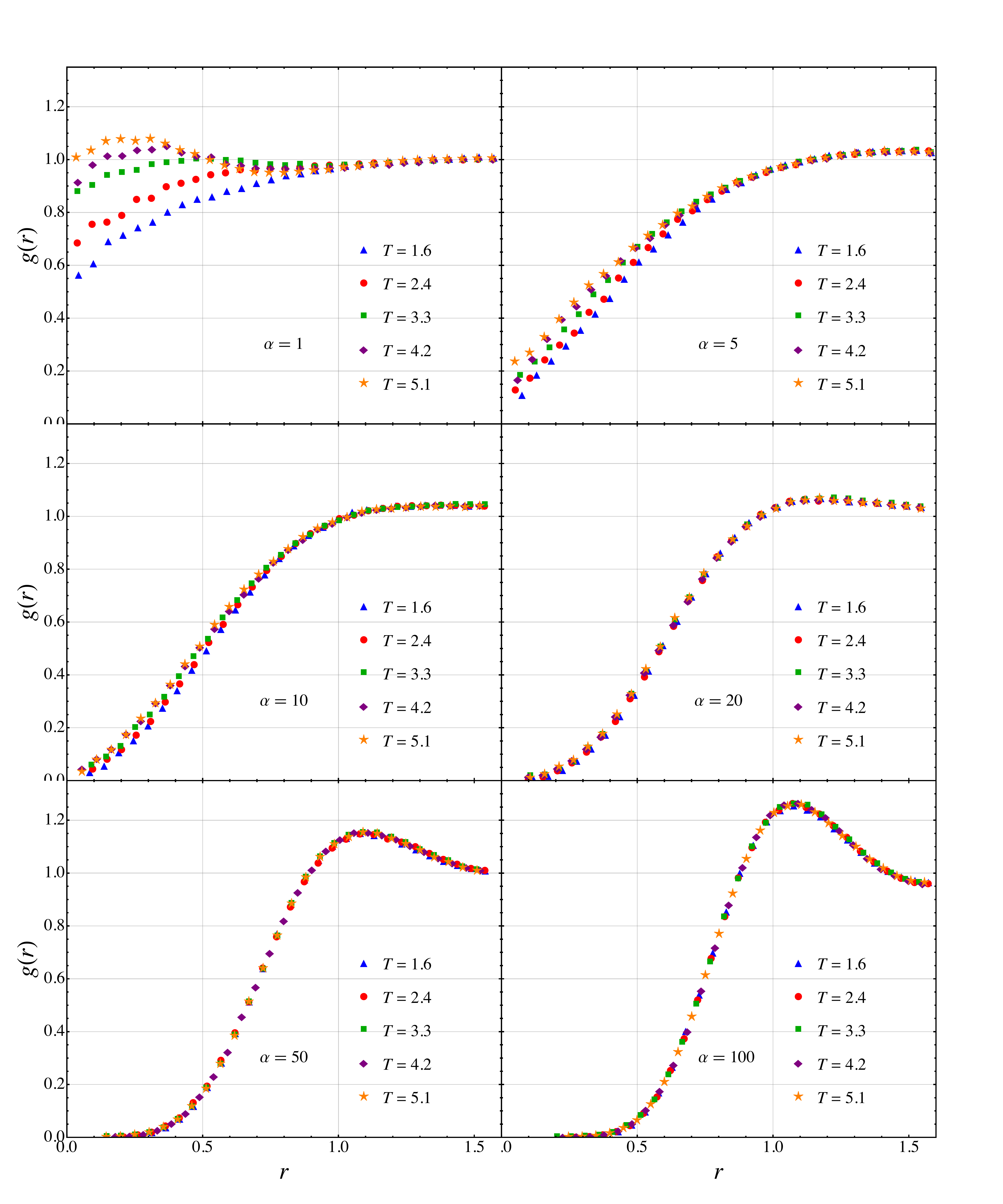}
        \captionof{figure}{Spatial correlations of the one-component Bose gas, at different temperatures and coupling strengths.}
        \label{fig_corr1}
\end{figure*}

Fig. \ref{fig_corr1} shows a sampling the radial correlations of the one-component Bose gas at different temperatures at couplings $\alpha =$ 1, 5, 10, 20, 50, and 100. 

We find that at weak couplings, $\alpha < 2$, the correlation functions flattened out as the temperature increased; this is caused by the fact that the thermal energy is greater than the potential repulsion in these cases. The slight increase in the correlation function near $r = 0$ is caused mostly by statistical fluctuations, but also in part by the fact that we are observing a jellium system with a neutralizing smeared background charge. At couplings $\alpha \geq 10$, the variations of temperature -- in the range we probed (0.5 $T_\text{c}$ to 1.8 $T_\text{c}$) are not reflected  in the correlation functions. As the coupling is increased the correlation functions show signs of structure, particularly at $\alpha \geq 20$.

\begin{figure*}
\begin{center}
\subfigure[]{%
\includegraphics[width=0.47\textwidth]{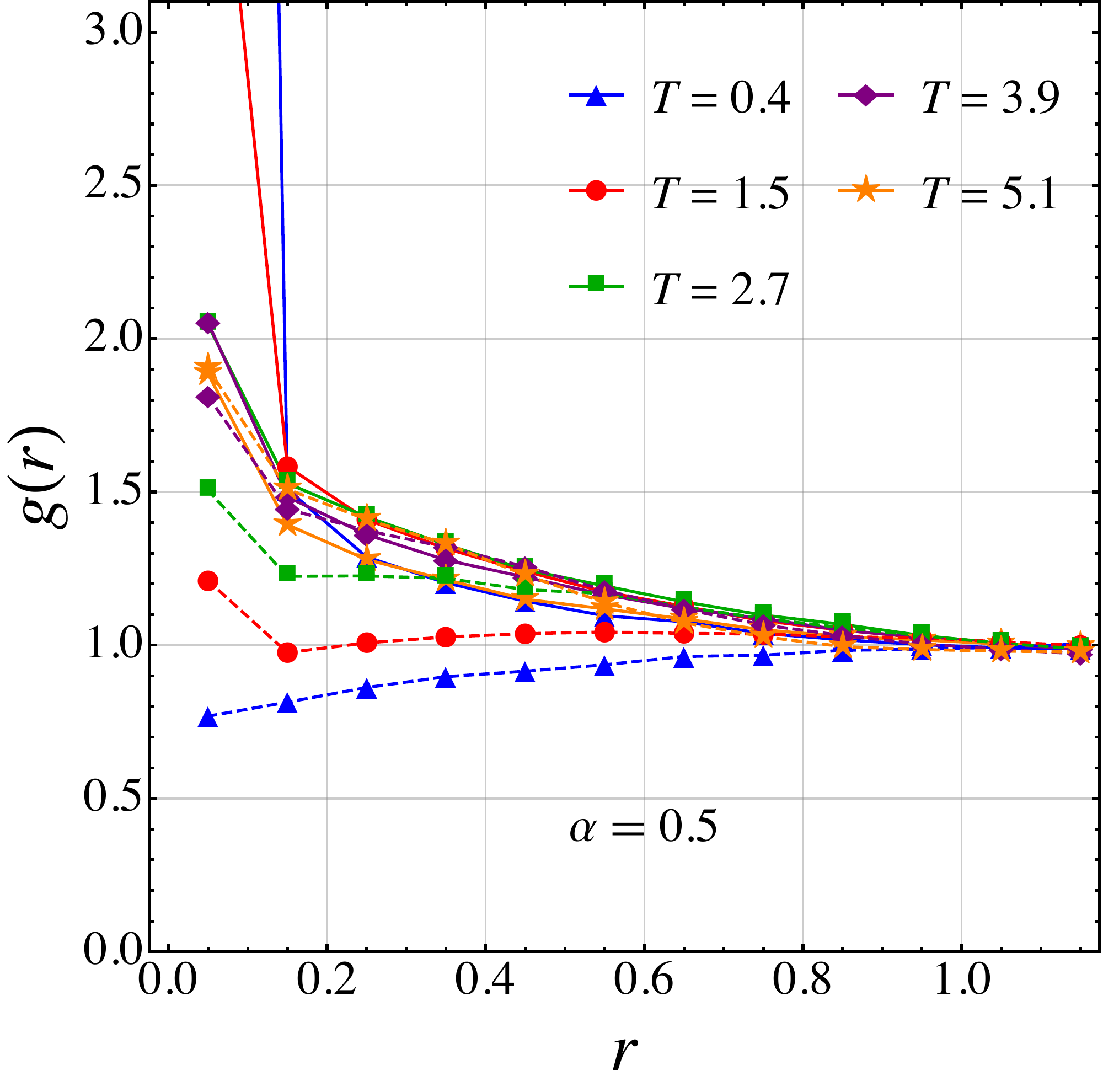}} \quad
\subfigure[]{
\includegraphics[width=0.47\textwidth]{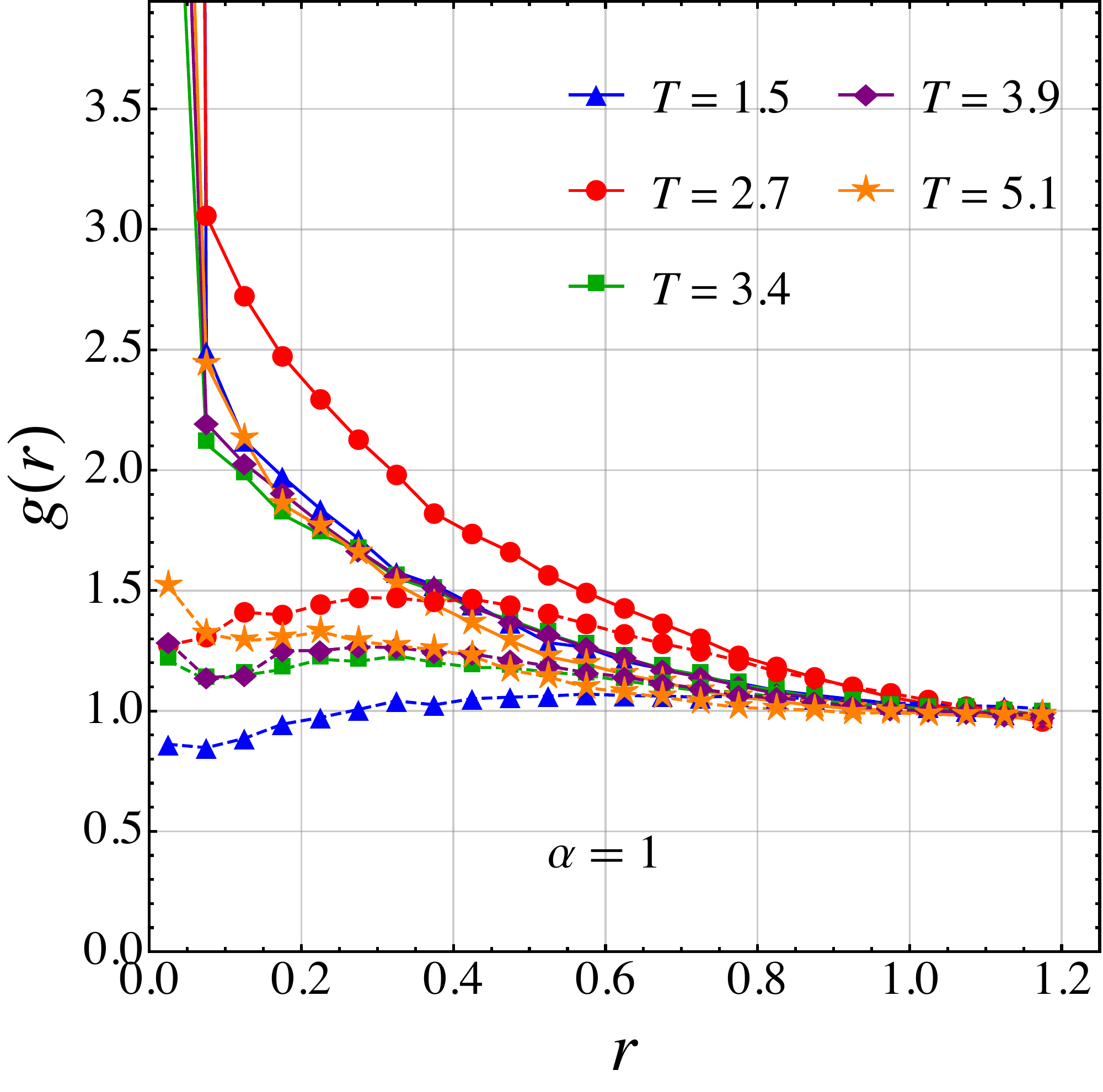}} \quad
\subfigure[]{
\includegraphics[width=0.47\textwidth]{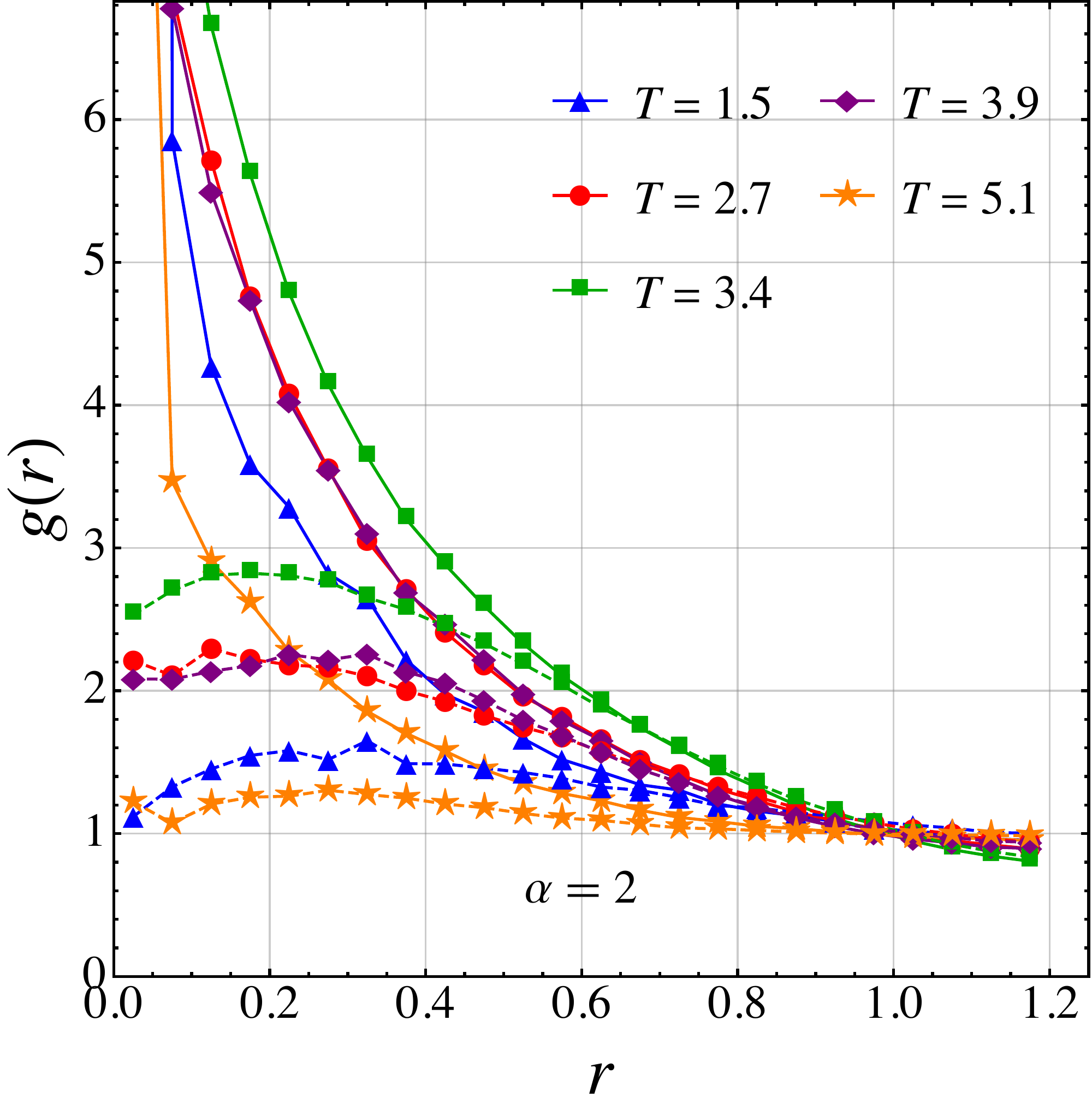}} \quad
\subfigure[]{
\includegraphics[width=0.47\textwidth]{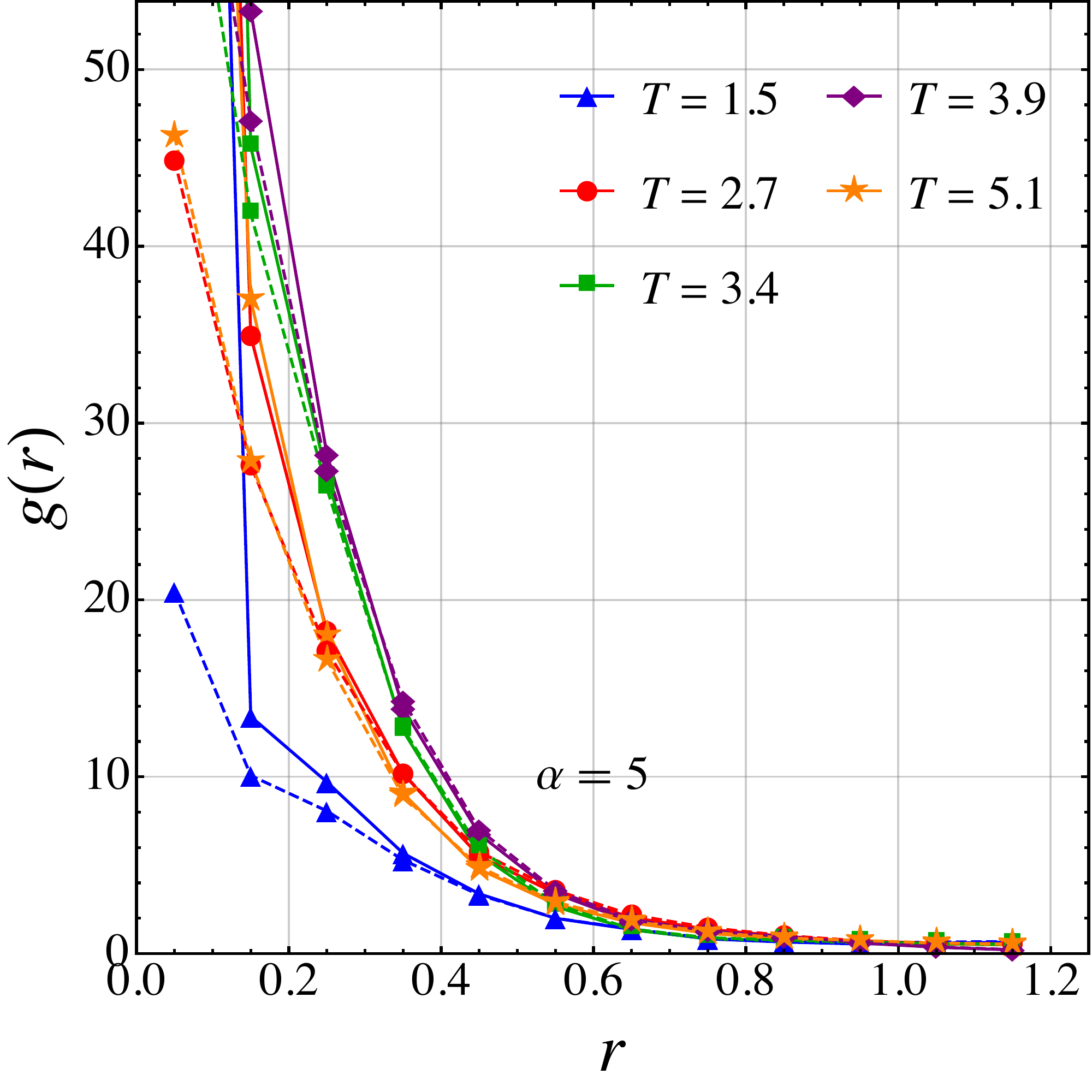}} \quad
\end{center}
\caption{Spatial correlations of the two-component Coulomb Bose gas (without core), at different temperatures and coupling strengths. Note that most plots have {\em two} correlation functions for each temperature, one ``attractive", for opposite sign charges $g_{+-}(r)$ (solid lines) and 
one ``repulsive", the same sign charges $g_{++}(r)$ (dashed lines). Note that at strong coupling, $\alpha \sim 5$, these two correlators  overlap significantly. }
\label{fig_corr2_1}
\end{figure*}

Fig. \ref{fig_corr2_1} shows the radial correlations of the for the two-component gas with no core at different temperatures at couplings $\alpha =$ 0.5, 1, 2, and 5. Without any core repulsions, at couplings $\alpha \geq 5$, the particles form small bound states; the particles essentially make point-like dipoles, especially at $T>T_\text{c}$. This is seen from the same-charge correlator being equal to the opposite-charge correlator at distances $r\geq 0.1$. An interesting feature seen in Fig. \ref{fig_corr2_1} is that both the anti-charge and same-charge correlations increase -- both in overall range as well as in magnitude at short range --  as $T$ approaches $T_\text{c}$ from below and then subsequently fall as temperature is increased further. The maximum short-range correlations occur slightly under $T_\text{c}$ for small couplings $\alpha \sim 1$, and move further above $T_\text{c}$ for larger couplings.

\begin{figure*}
\includegraphics[width=\textwidth]{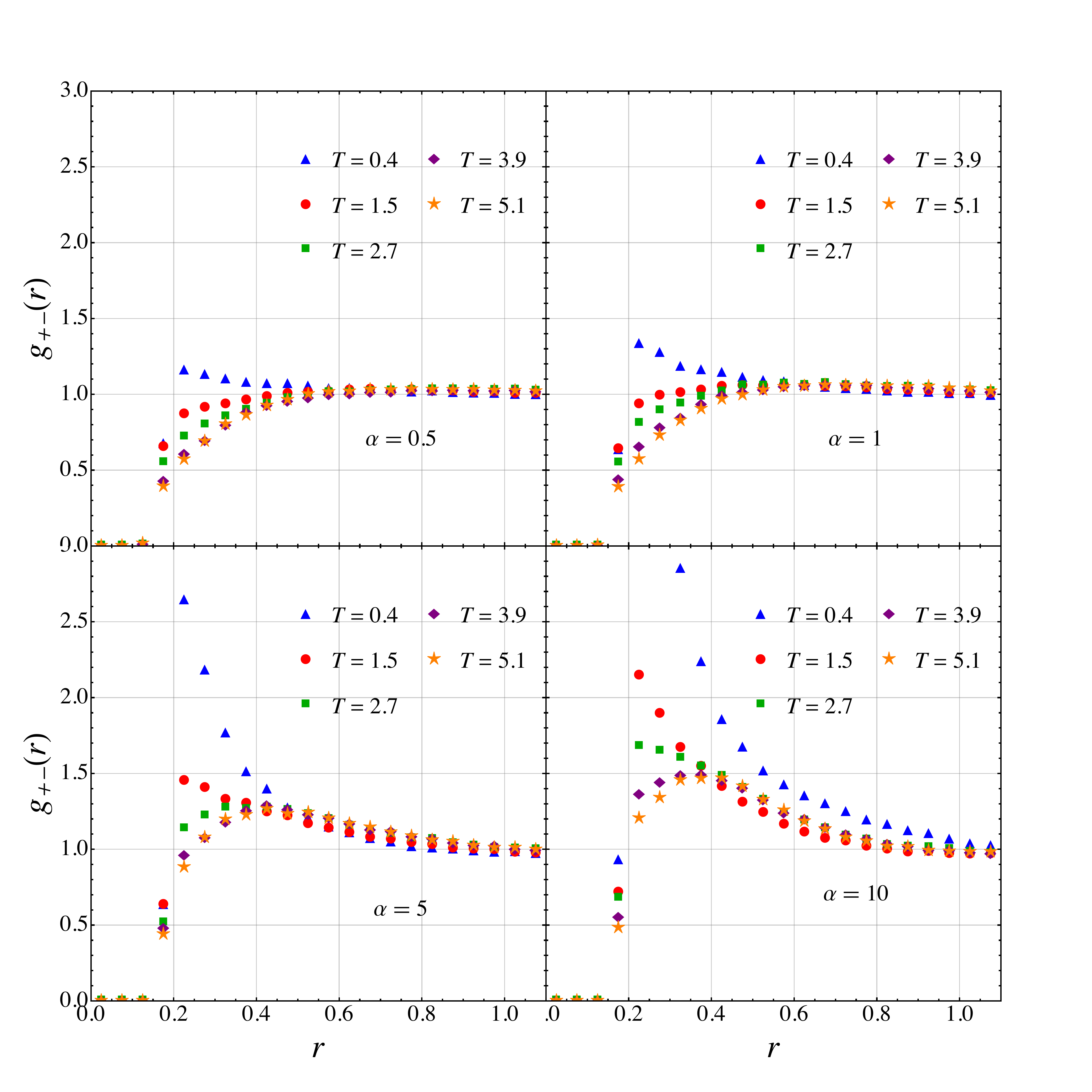}
\caption{Spatial correlations of the two-component Bose gas with core repulsion for particles of opposite charge, at different temperatures and coupling strengths.}
\label{fig_corr2_2pm}
\end{figure*}

\begin{figure*}
\includegraphics[width=\textwidth]{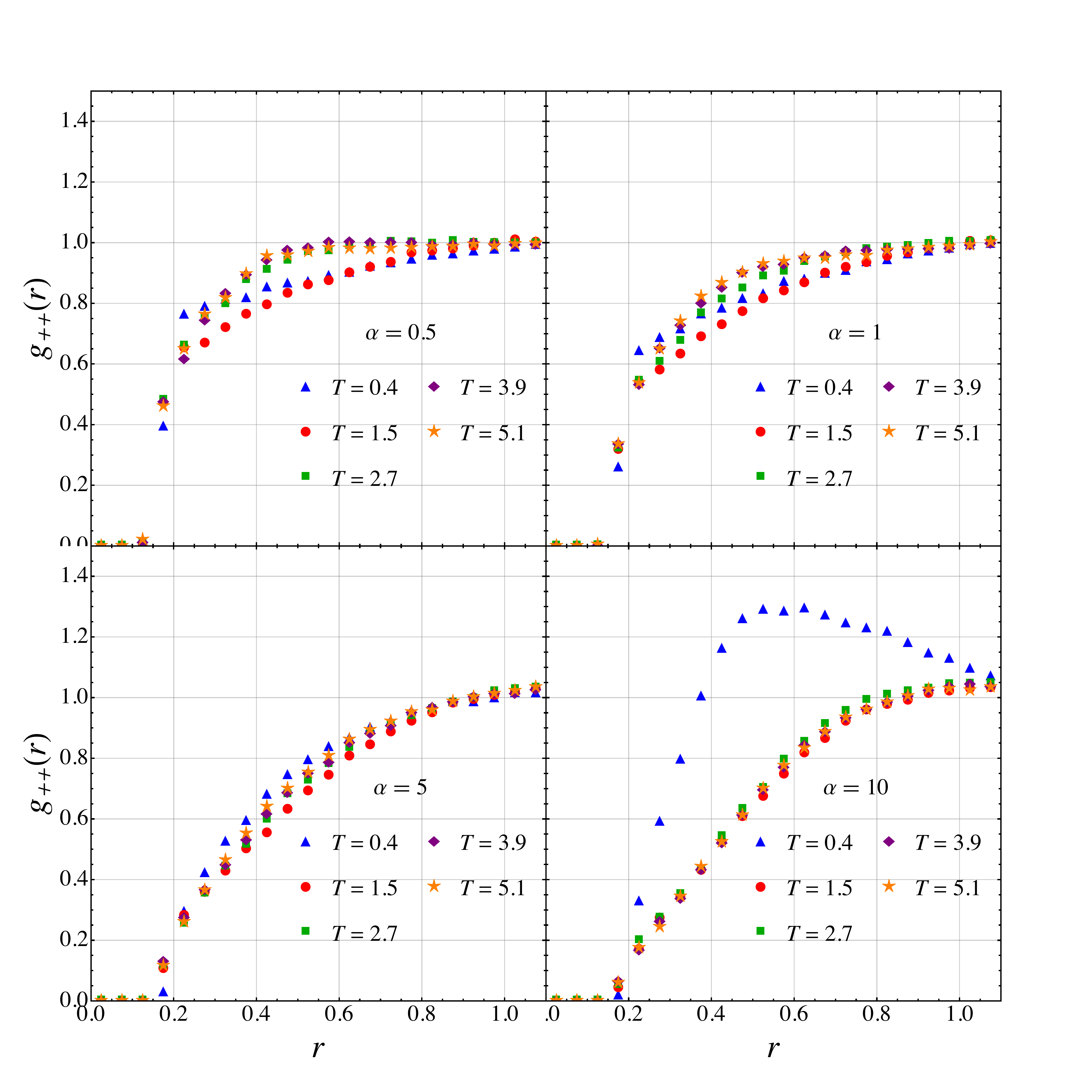}
\caption{Spatial correlations of the two-component Bose gas with core repulsion for particles of the same charges, at different temperatures and coupling strengths.}
\label{fig_corr2_2s}
\end{figure*}

Figs. \ref{fig_corr2_2pm} and \ref{fig_corr2_2s} show the radial correlations of the for the two-component gas with a repulsive core at different temperatures at couplings $\alpha =$ 0.5, 1, 5, and 10. At low temperatures, $T<T_\text{c}/3$, the probability for there to be two oppositely charged particles in a bound state is large even at small couplings $\alpha < 1$; at larger couplings, we find bound states at higher temperatures. At $\alpha \sim \mathcal{O}(10)$ and larger, at low temperatures, we see the same screening phenomenon we saw when there was no repulsive core; the molecule acts as a neutral dipole, which causes the same-charge correlator (Fig. \ref{fig_corr2_2s}, bottom right panel) to be {\em greater} than unity.

\subsection{Thermodynamics}\label{tcthermo}

\begin{figure}[h!]
\begin{center}
\includegraphics[width=\columnwidth,angle=0]{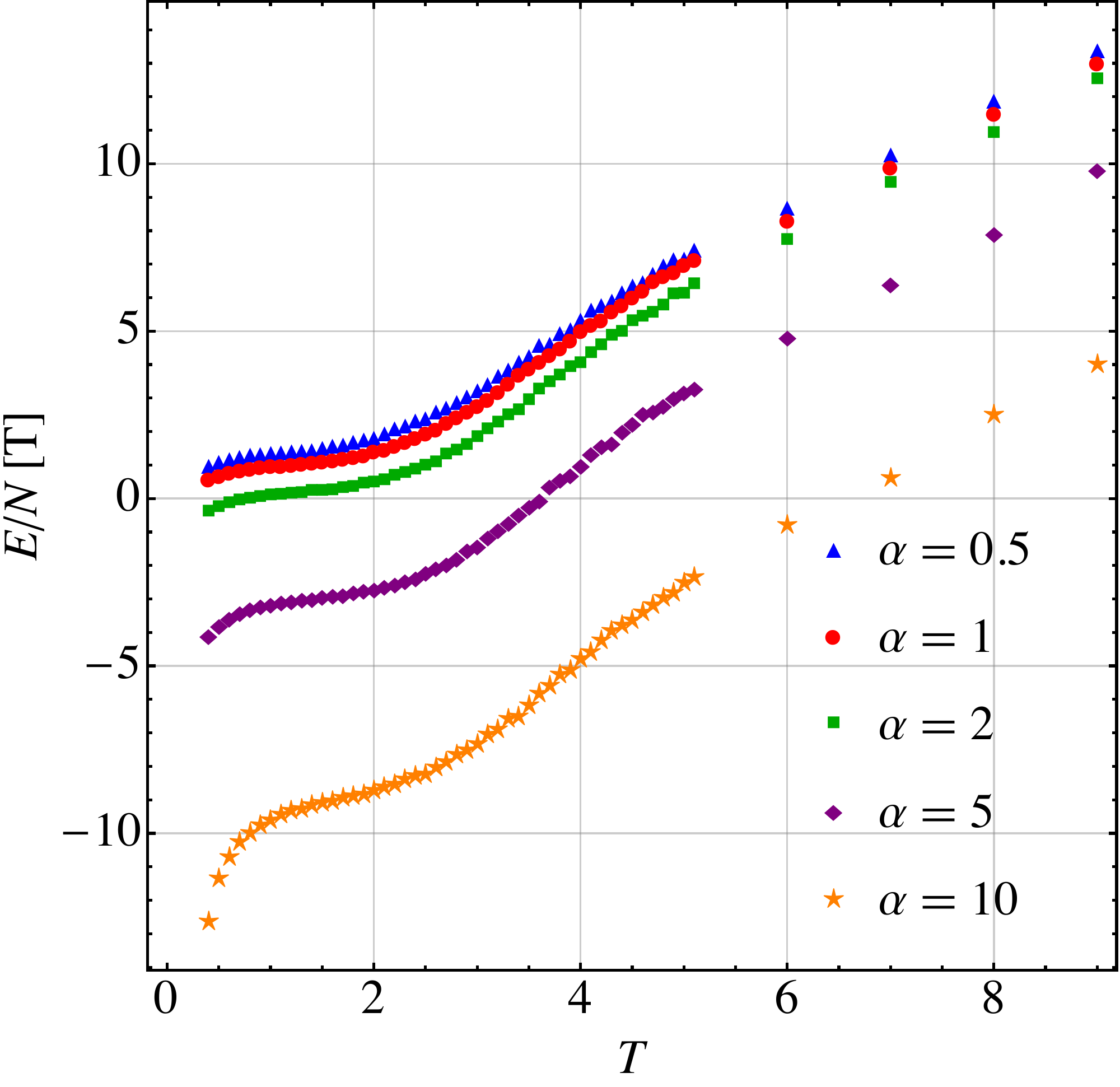}
\caption{Internal energy per particle of the two-component Coulomb Bose gas with a repulsive core, for various couplings and temperatures. Error bars are smaller than the points themselves.}
\label{fig_en}
\end{center}
\end{figure}

Fig. \ref{fig_en} shows the internal energy per particle of the two component gas with a core, in units of the temperature, across various temperatures and couplings. Let us remind the reader that in the 2-body Coulomb problem, the virial theorem tells us that mean potential energy is $-\sfrac{1}{2}$ times the mean kinetic energy, so the total energy is positive. Many-body strong coupling problems, on the other hand, can create crystal-like correlations between many particles, producing  larger potential energy, and thus  negative total energy.

The temperature dependence of the energy is similar for all couplings, and, at fixed temperature, the energy scales roughly linear with temperature. At high temperatures, the kinetic energy begins to scale at $\sfrac{3}{2}T$, as predicted by classical statistical mechanics. One can see that, at high coupling, the energy falls rapidly as temperature is decreased to near zero, which reflects the increasing binding of the oppositely charged particles. 

\section{Effective Model of Color Magnetic Monopoles}\label{eff}

Classical studies of the magnetic scenario in QGP proposed that the magnetic component of the plasma acts as a liquid with Coulomb-like fields \cite{Liao:2006ry, PhysRevLett.101.162302}, and contemporary studies on the lattice, e.g  \cite{DAlessandro:2007lae, D'Alessandro:2010xg}, furthered the study of monopoles in QCD-like theories. These studies on the lattice \cite{DAlessandro:2007lae} found that the monopole density at $T>T_\text{c}$ can be well approximated by 
\eq{
\rho_m(T) \sim {T^3 \over \log{T}^{2}}\,.
}
Unlike the ``electric" particles, quarks and gluons, the density of monopoles is not vanishing at $T_\text{c}$ due to confinement, but instead has a {\em peak} there. It also follows from the correlation function analysis that the magnetic coupling becomes stronger as temperature increases \cite{PhysRevLett.101.162302}.  

Based on these findings, we would like to make an effective model of quantum monopoles that reproduces the behavior of those on the lattice, without the many degrees of freedom of a full QCD-like theory.

\subsection{Correlation Function Matching}\label{match}

\begin{figure*}
\includegraphics[width=\textwidth]{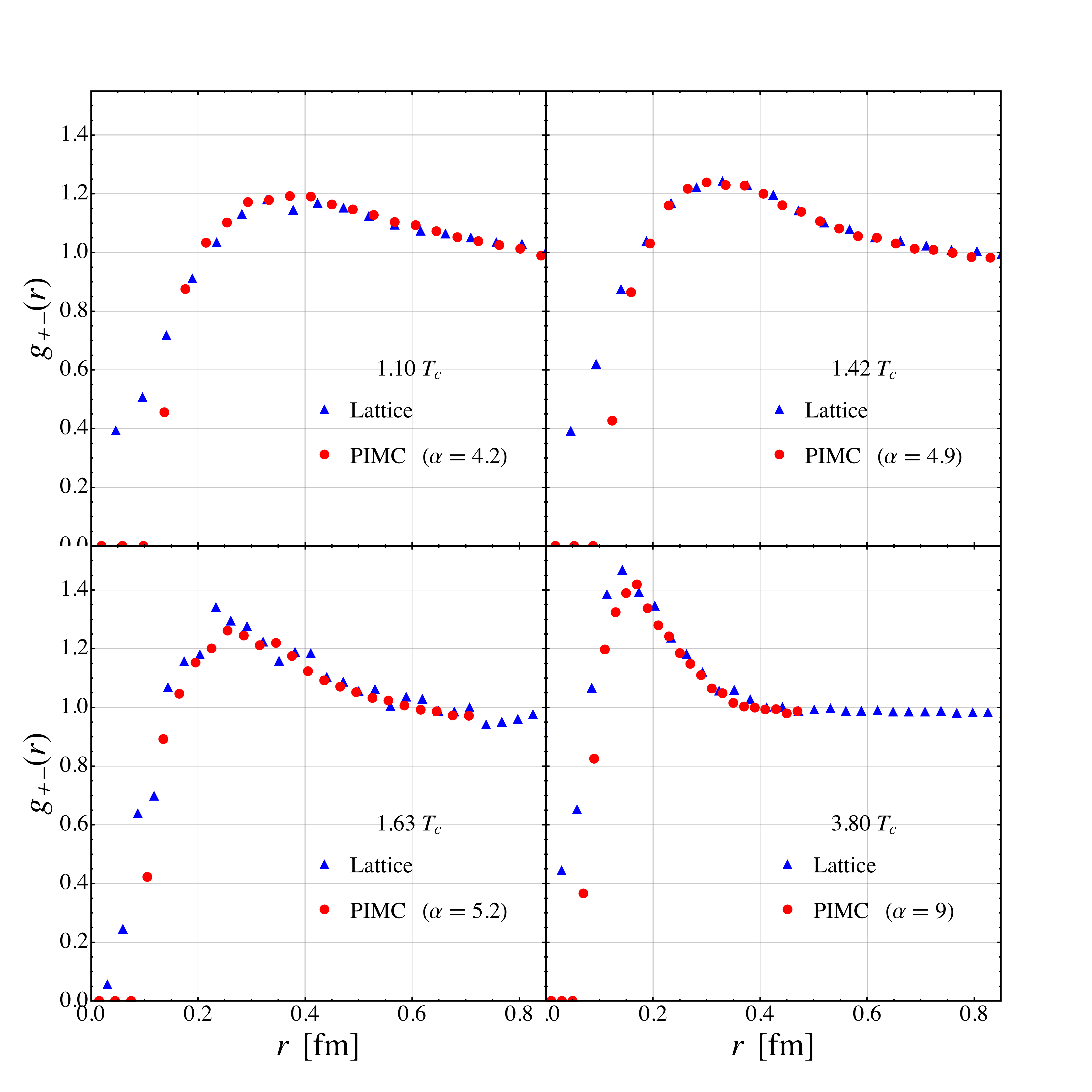}
\caption{Spatial correlations from our simulations (red circles) matched via scaling to lattice correlations (blue triangles) from  \cite{DAlessandro:2007lae} at various temperatures.}
\label{fig_latcorr}
\end{figure*}

Before we dig out into details of the matching procedure, let us outline its general meaning. The simulated Coulomb Bose gas  model has several parameters, such as the density $n$, the temperature $T$, the Coulomb coupling $\alpha$, and the particle mass $m$.  The monopole ensemble corresponding to the finite temperature QCD has only one input variable, $T$, and thus only simulations on a particular parametric line $n(T),\alpha(T), m(T)$  in the general parameter space are directly relevant for our physics application. 

Now that we have quantified the behavior and thermodynamics of an isolated two-component Coulomb Bose system, our first goal is to find the parameters for our model that are necessary to effectively model magnetic monopoles in QCD-like theories at various temperatures above the critical temperature. To fit our findings to physical results, we first compare our correlation functions with those of \cite{DAlessandro:2007lae}, found on the lattice. We note that this lattice calculation was done in pure-gauge $SU(2)$, which yields one $U(1)$ monopole species. 

The lattice correlations and the matching correlation functions from our simulations are seen in Fig. \ref{fig_latcorr}. We match these two sets of correlation functions by scaling our inter-particle distance to that given by the monopole density in \cite{DAlessandro:2007lae}, and then by finding the simulation coupling strength that produces the same magnitude and long-range correlation behavior seen on the lattice.

First and foremost, we see that a two-component Coulomb Bose system reproduces the same types of correlations seen on the lattice, as was found in \cite{PhysRevLett.101.162302}, giving further credence that our model can effectively describe the behavior of magnetic monopoles in QCD-like theories. The mapping of our results to those on the lattice is given by
\eq{\alpha(T) \approx 3.4\, \rho_m^{1/3}(T)\,,} 
where $\alpha(T)$ the coupling used in our simulation, $\rho_m(T)$ the monopole density (in fm$^{-3}$) found in \cite{DAlessandro:2007lae}, and $T$ in units of the critical temperature. One unit of length in our simulations is equivalent to $\rho_m^{-1/3}(T)$, the interparticle spacing (in fm) found on the lattice. This result was checked in and holds throughout the range 1.1-4$T_\text{c}$; it may be applicable at higher temperatures as well. The constant 3.4 is comes from the factor of $(T_\text{c})^3$, which sets the scale of the density in dimensionless units.

If this relation holds to $T_\text{c}$, we can map the physical value of the critical temperature to our units: 296 MeV ($T_\text{c}$ in the $SU(2)$ lattice simulation \cite{DAlessandro:2007lae}) is the critical temperature of the two-component gas with coupling $\alpha\approx4.2$ -- approximately 3.45 in our units. We found that a two-component Coulomb Bose gas with $\alpha\approx4.2$ has a 5-10\% higher critical temperature than the free gas; the critical temperature of a free gas of monopoles would then roughly correspond to 280 MeV. Extrapolating the monopole density from \cite{DAlessandro:2007lae}  to $T_\text{c}$, setting $T_\text{c}$ to 280 MeV, and solving for mass in Einstein's equation, we find that
\eq{
280\text{ MeV} =    \left(\frac{2\pi}{m}\right)\left( \frac{(240 \text{ MeV})^3}{\zeta\left(\frac{3}{2}\right)}\right)^\frac{2}{3} \rightarrow m \approx 680 \text{ MeV}\,,
}
which is approximately the estimate of the monopole mass from \cite{D'Alessandro:2010xg}, though it is larger than that from \cite{Cristoforetti:2009tx}.

\subsection{Monopole Contribution to $SU(2)$ Thermodynamics}

After the parameters of our model were matched to those in the pure-gauge $SU(2)$ lattice simulations, we can directly evaluate the contribution of the monopoles to the thermodynamics of that theory. 
  
   \begin{figure}
\begin{center}
\includegraphics[width=.5\textwidth,angle=0]{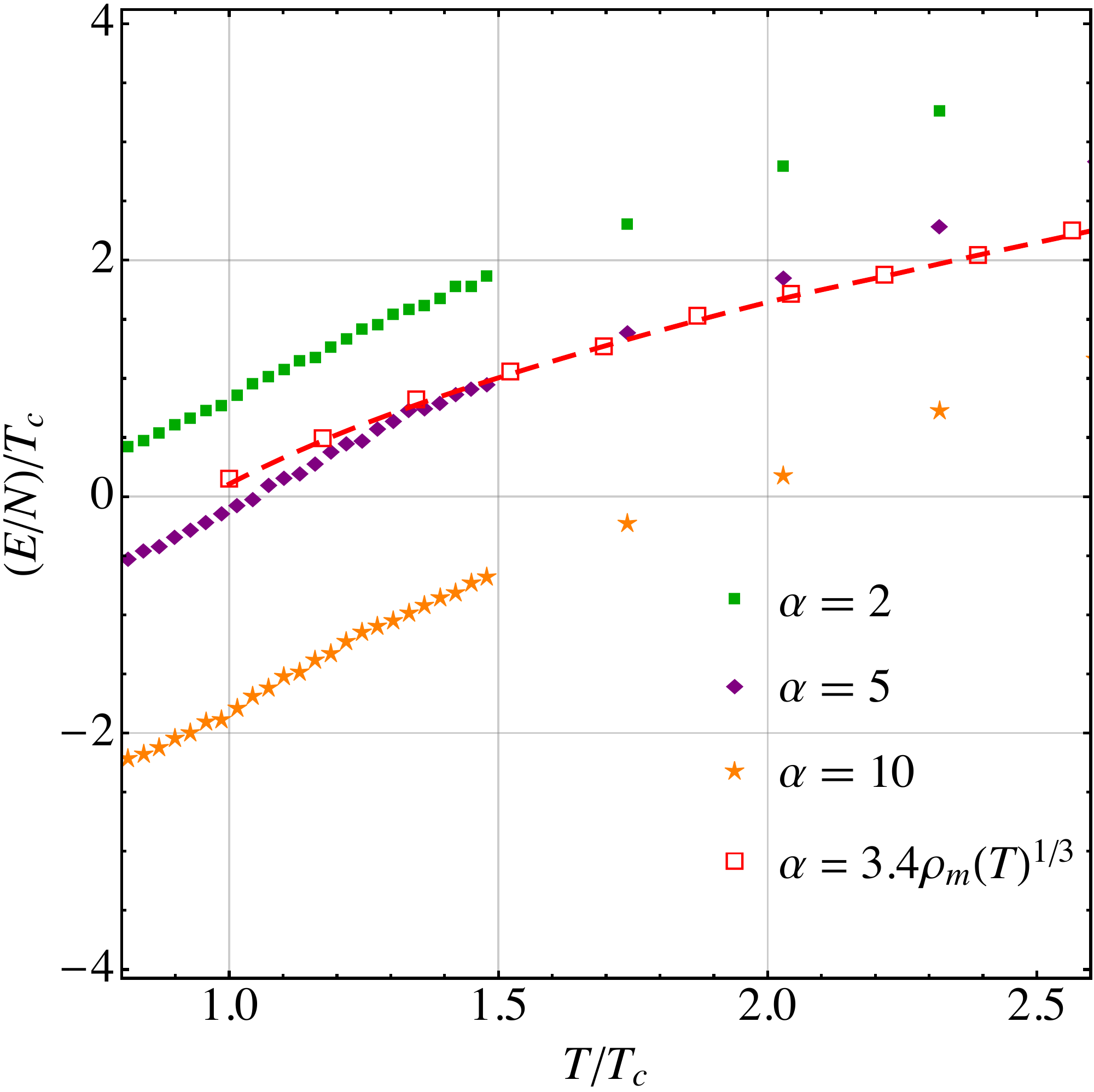}
\caption{The energy per particle along the physical line defined by the parameters which match simulation correlation functions to the lattice, shown alongside previously shown two-component Coulomb simulations (Fig. \ref{fig_en}) at fixed couplings. The dashed line is to guide the eye.}
\label{fig_physline}
\end{center}
\end{figure}

   \begin{figure}
\begin{center}
\includegraphics[width=.5\textwidth,angle=0]{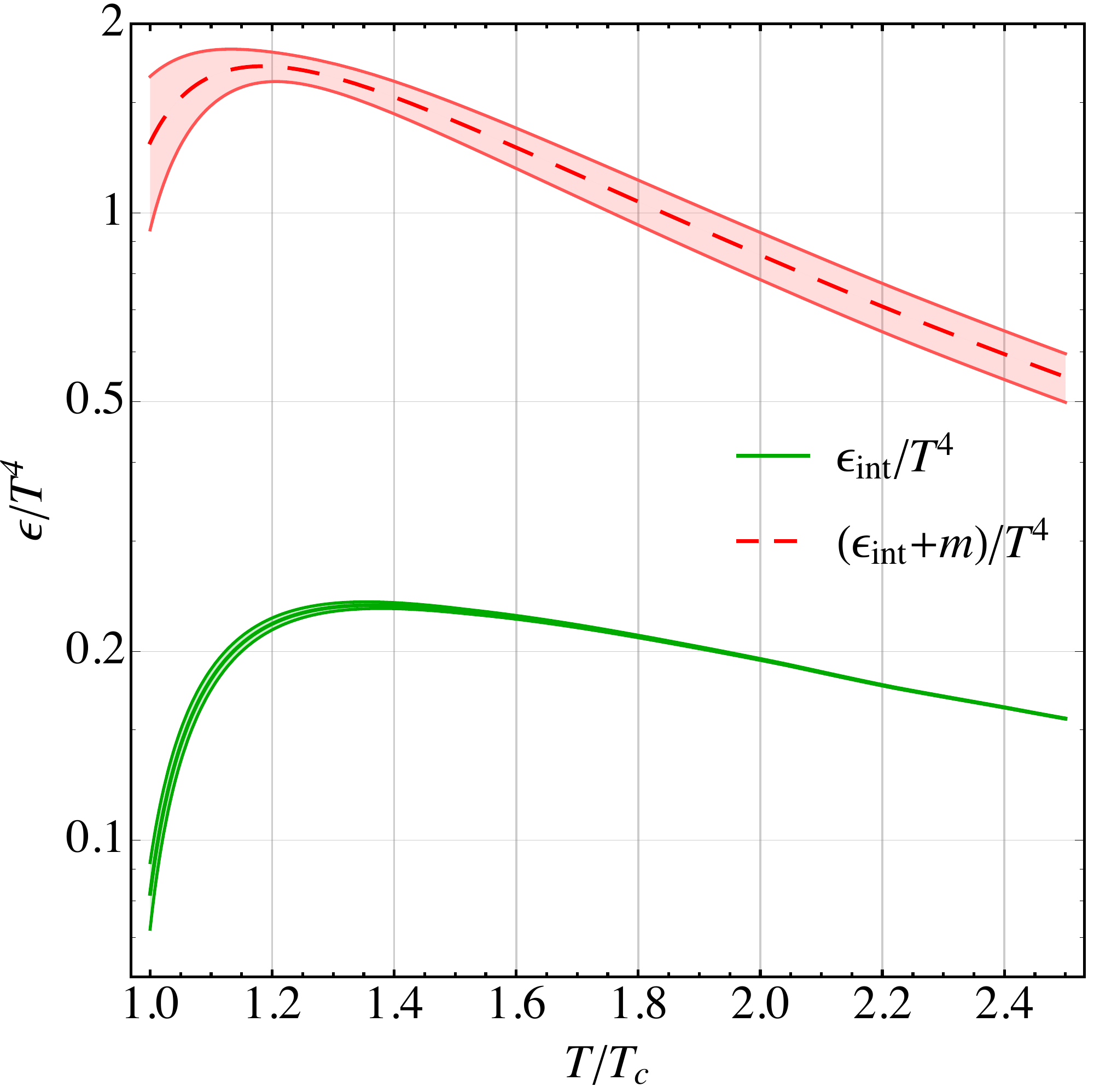}
\caption{The dimensionless energy density for $SU(2)$ monopoles along the physical line defined by the parameters which match simulation correlation functions to the lattice. The lower line is the internal energy density (kinetic + potential) and the upper line includes the mass contribution.}
\label{fig_su2en}
\end{center}
\end{figure}

The thermodynamics on the {\em physical line}, i.e. the trajectory in parameter space as defined above in Sec. \ref{match}, are shown in Figs. \ref{fig_physline} and \ref{fig_su2en}. In Fig. \ref{fig_physline}, we see the energy per particle along the physical line. At $T_\text{c}$, the internal energy of a monopole goes to approximately zero, and then grows as temperature rises. This growth is less than $\sfrac{3}{2}T$ because the coupling is increased with $T$, causing monopoles at large $T$ to have a larger negative potential energy. This lower energy is, however, compensated by the $\sim T^3/(\log{T})^2$ growth of the density of monopoles. 

In Fig. \ref{fig_su2en}, on the lower solid line, we see that the internal energy density of the monopoles $\epsilon/T^4$ is maximal at approximately 1.3-1.4$T_\text{c}$. The contribution from the monopoles is relatively small, $\mathcal{O}(0.2)$.

In light of the fact that this contribution is close to zero, in addition to the internal energy (kinetic and potential), we must also take into account the mass of the particles, which has been extracted from the lattice in \cite{D'Alessandro:2010xg}. In Fig. \ref{fig_su2en}, the upper dashed line shows the total energy density of the monopoles. The contribution from the monopoles, including the mass, is $\mathcal{O}(1.5)$.

\subsection{Generalization to the Thermodynamics of  QCD} 

QCD thermodynamical quantities are among the most basic properties of hadronic matter, and have been the focus of numerous lattice simulations for the last three decades. Due to growth of computational power and algorithm development, the results of these calculations have become rather accurate, and, over the last few years, have approached QCD with physical quark masses. We now know the pressure (free energy), energy, and entropy density as a function of the temperature: $p(T)$, $\epsilon(T)$, and $s(T)$, respectively.

At first glance, decomposing those functions into parts associated with certain quasiparticles -- gluons, quarks, and perhaps monopoles --  looks to be an impossible task, since all of them interact strongly. However, more recent studies have provided valuable insight, indicating that this task can perhaps be carried out.

The first step toward understanding of the role of monopoles in QCD is to move from the $SU(2)$ to the $SU(3)$ color group.
This is far from trivial, since the latter has two diagonal color generators, and thus two ``massless electrodynamics" surviving the breaking of the color group. Therefore, there are two distinct species of the monopoles. Including the anti-monopoles, one would need to study a {\em four}-component Coulomb Bose gas. This system can be studied in the same way as for the one and two-component gases above, but was not done for this work. Note that two species of monopoles are not independent, as there should be attractive Coulomb forces between the monopoles of each $U(1)$ electrodynamics, as well as repulsive forces between monopoles and anti-monopoles of different $U(1)$ electrodynamics.
 
A comprehensive study of the condensation and density of lattice monopoles for  the $SU(3)$ color group has been done by Bonati and D'Elia \cite{Bonati:2013bga}. As shown in Fig. 3 of that work, both monopole species happen to have nearly identical densities. Moreover, Fig. 2 and the corresponding text from \cite{Bonati:2013bga} indicate that, for each of them, the density $\rho(T)$ is very close to that of the $SU(2)$ monopoles; these densities effectively identical when taking into account the difference in $T_\text{c}$ for the different $SU(2)$ and $SU(3)$ simulations. However, they have a slightly different fit from the $SU(2)$ data at high $T>2T_\text{c}$, namely
\eq{  \rho_m(T)={3.66 T^3 \over \log({T/(0.163 T_\text{c})})^{3}}\,.
}
This power of the logarithm matches predictions from the 1970s for finite-$T$ QCD: the magnetic scale cubed  $(g^2 T)^3$.

The next step toward QCD would be to include quarks. The corresponding lattice simulations are unfortunately very expensive, especially for  quarks as light as those in the real world. Only relatively recently have such lattice ensembles became available, and the analysis of their monopole content  has not yet been done. Lacking simulation input, we will  provide some speculation on ``theoretical expectations."

Let us start from the high-$T$ end, simply counting states. Gluons have two polarizations and $N_c^2-1$ colors, giving 16 bosonic states.  Quarks contribute  $2\cdot 2\cdot N_c N_f\cdot (7/8)$ (for $N_c=N_f=3$, $36\cdot(7/8) =31.5$) times the thermal energy of one bosonic state. Monopoles are charged spin-zero scalars, of $2(N_c-1)$ types (the number of diagonal generators multiplied by 2 to take into account the two charges), or 4 species for $N_c=3$. At high $T$ they only exist at the so-called magnetic scale, and thus their density is additionally suppressed by a power of $\log(T)$, as discussed above. As a result, at high $T$, the monopole contribution is quite small as compared to that of quarks and gluons.

This is not the end of the story, however, because the light quarks can be bound to the monopoles. The corresponding Dirac equation has no coupling, and the fermionic 3d zero modes are of topological nature, thus they should be present at {\em any} $T$. While these bound states are scalars -- the spin 1/2 and the color spin 1/2 add up to the grand spin 0 -- the quark zero mode can be either occupied or empty, interpreted as 2 separate states, a doublet of the baryon number  $B=\pm 1/2$ \cite{Jackiw:1975fn}. 
So, in the theory with a single light quark,
$N_f=1$, the number of magnetic states doubles.

 In a theory with $N_f=2$, one can get the $B=1$ {\em triplet} of states. Its flavor-asymmetric wave function can be, for example, viewed as the isospin-0 $ud$ diquark, the antidiquark, and also the $\eta$ isoscalar meson added to a monopole. The number of magnetically charged states is, in this case, $2\cdot 2\cdot 3=12$. While this is still smaller than the number of quark and gluon degrees of freedom, it is not a negligible contribution.

A qualitative observation made by Liao and Shuryak \cite{Liao:2012tw} was that, with the number of monopole-quark species growing with $N_f$, it becomes more and more difficult to produce Bose-Einstein condensation, since the objects become distinguishable. This tendency can only be counterbalanced by a corresponding increase of the monopole density. And indeed, lattice simulations for QCD-like theories with an increasing $N_f$ have found that deconfinement transition corresponds to stronger  coupling $g^2(T_c)$, smaller  monopole mass and therefore higher monopole density.

All dimensional quantities
are defined following standard lattice convention for units:  the
vacuum string tension for all theories is declared to be the same in MeV.
With such units, the critical temperature for   $SU(2)$ and $SU(3)$ is different,
 $\sim300$ MeV for $SU(2)$ and $\sim260$ MeV for $SU(3)$, but the densities of each of the $SU(2)$ and $SU(3)$ monopoles are about the same \cite{Bonati:2013bga}. If $T_\text{c}$ is lower, the overall density of monopoles grows, so the density of each separate species of monopole becomes large enough to form a  Bose-Einstein condensate. Recall that, as was found in \cite{Bonati:2013bga}, we observed that the inclusion of an additional interacting component to a Bose Coulomb system did not alter the critical temperature behavior, provided the density of each component was not altered.

The spectroscopy of quark-monopole states in QCD at zero temperature would be very hard to study, because the hybrid (meson-glueball) states would be heavy and wide, mixing with many other mesonic states. But at $T\approx T_\text{c}$, where the monopoles are relatively light, these states can perhaps be identified. Theoretically, it is also hard to predict their masses; while the Dirac equation for quark fields are indeed written exactly without any coupling present, the Yang-Mills equations for the monopole gauge field itself has only been solved in the classical approximation, in which it is assumed that the monopole action is much larger than the (one-loop) quark correction to it. As discussed above, this is no longer so near $T_\text{c}$.

Going back to the thermodynamics of QCD, a historically important argument has been related with the mean value of the  Polyakov line, $\braket{P(T)}$. According to lattice data,  $\braket{P(T)}$ approaches 1 at high $T$ very slowly: it reaches 0.8 at $T\approx 350$ MeV according to Fig. 3 of \cite{Borsanyi:2010bp}, or  $T\approx 500$ MeV according to Fig. 1 of \cite{Petreczky:2015yta}. This quite far from $T_\text{c}\approx 155$ MeV.  
  
While literally it should be applied to the static quarks, one can $conjecture$ that the thermodynamical contributions of light quarks should also be proportional to it,
 \eq{n_q(T)\sim \braket{P(T)}\,.}
This led to the development of the Polyakov-Nambu-Jona-Lasinio (PNJL) model \cite{Meisinger:1995ih,Fukushima:2003fw} and similar models. Direct lattice studies, e.g. \cite{Bazavov:2013dta}, were able to identify the density of strange quarks, $n_s(T)$, using a certain combination of susceptibilities vanishing for mesons and baryons, but not for quarks. Their results confirm this  conjecture rather well. One may further argue that the density of (color non-diagonal) gluons should  then be proportional to the square of the Polyakov line,  $\braket{P(T)}^2$. If so, $n_g(T)$ must be even more suppressed near $T_\text{c}$ than $n_q(T)$.   
  
At the same time, the energy and entropy densities, $\epsilon/T^4$ and $s/T^3$, respectively, rise to their approximate scale-invariant value much more rapidly, by $T-T_\text{c}\sim50$ MeV or so, unlike 200-300 MeV for  $\braket{P(T)}$. See, for example, Fig. 6 of \cite{Bazavov:2014pvz}. The inevitable conclusion from these arguments is that there must be some extra contribution, in addition to the quarks and gluons, in this interval of temperature. 
  
It was suggested, e.g. in \cite{Shuryak:2004tx}, that there should be {\em bound states} of quarks -- mesons and baryons -- at $T>T_\text{c}$. The presence of those  are indeed well documented now on the lattice, e.g. in \cite{Bazavov:2013dta} for strange quarks and later for charmed ones.
  
We now turn to the following question: what is the contribution of the {\em monopoles} to the global thermodynamics? Since the monopoles are identified on the lattice individually, with their Euclidean-time paths and correlation functions determined from simulation, it should be possible to calculate their energy. Lamentably, this has not been done yet.

\begin{figure}[h!]
\begin{center}
\includegraphics[width=.47\textwidth,angle=0]{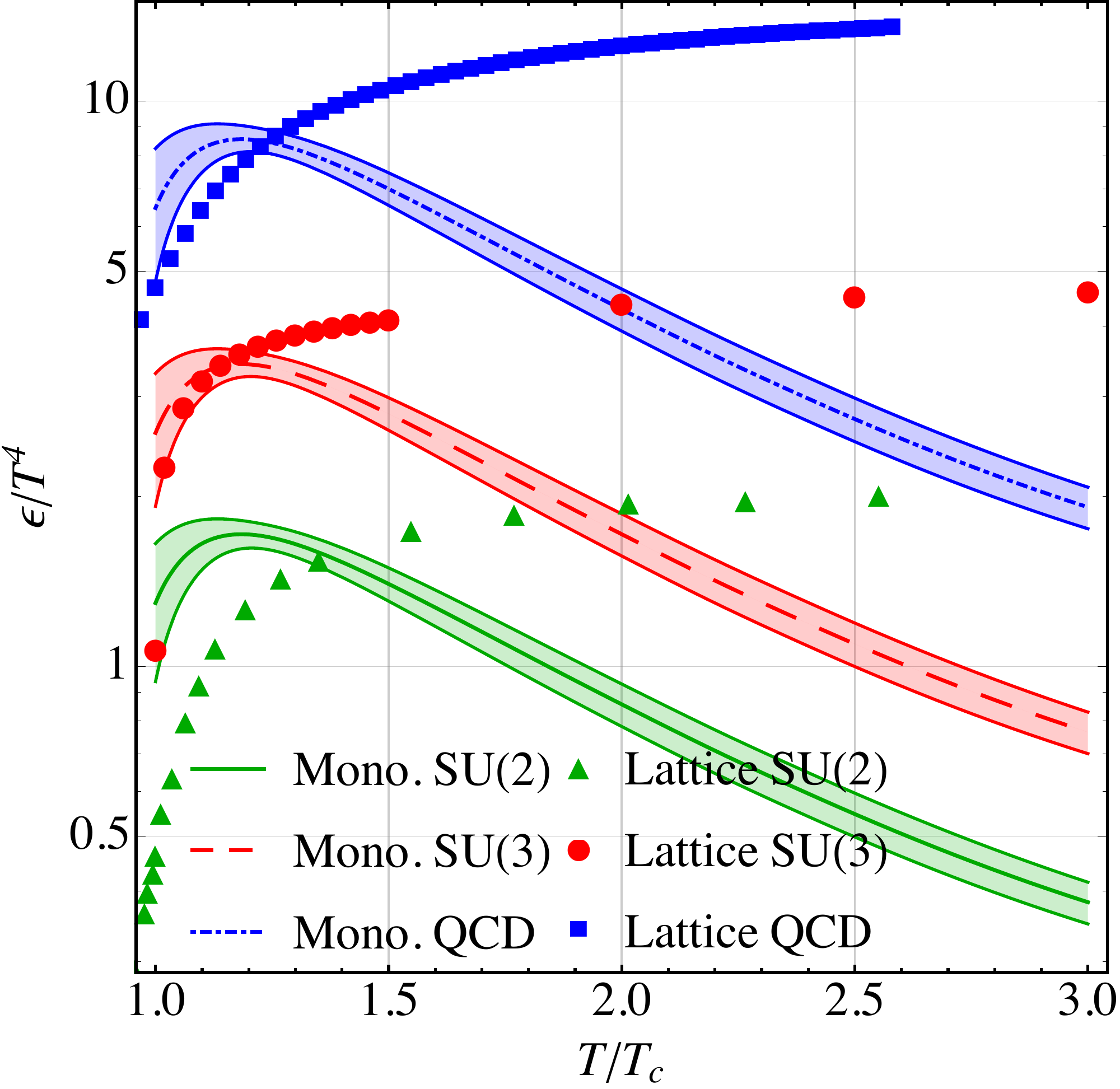}
\caption{The energy density of the monopoles compared to lattice data for pure-gauge $SU(2)$ and $SU(3)$. The estimates for the monopole contribution to $SU(3)$ and QCD are from a scaling argument (see text). Lattice $SU(2)$ results are from \cite{Engels:1994xj,Engels:1990vr}, $SU(3)$ from \cite{Borsanyi:2012ve}, and QCD from \cite{Bazavov:2014pvz}.}
\label{fig_latencomp}
\end{center}
\end{figure}

A comparison of the energy density contribution of monopoles to the overall energy density found on the lattice is seen in Fig. \ref{fig_latencomp}. The $SU(2)$ comparison is direct from our study, as this is the same system from which we found our parameter fits. For $SU(3)$ on the other hand, we have estimated the monopole contribution to the energy density by simply multiplying the contribution by 2, as discussed above. This, of course, does not take into account the energy coming from the interaction between the two species, which will have to be studied further in the future. Finally, for QCD, from standard counting of degrees of freedom, one finds that quarks have about twice more of those than gluons, so the QGP energy density in QCD is about 3 times larger than in the pure gauge $SU(3)$. The number of quark-monopole states, as we argued above, for two light flavors also increase the number of species by the factor of 3.

The monopole mass, as found in \cite{D'Alessandro:2010xg}, contributes significantly more to the overall thermodynamics than the internal energy of the monopoles  (c.f. Fig. \ref{fig_su2en}). In Fig. \ref{fig_latencomp}, we see the total energy density contribution from the monopoles to the system. When the mass is included, the monopoles contribution constitutes the {\em entire} energy density of the system between 1-1.3$T_\text{c}$.
   
\section{Conclusions and Outlook}

In this work, we have studied the effects of Coulombic interaction on the Bose-Einstein condensation. We numerically calculated the critical temperature $T_\text{c}$, of one- and two-component Coulomb systems, by two different methods, as a function of the interaction strength. Qualitatively,  the dependence is the similar to what has been previously observed for hard spheres: $T_\text{c}$ moderately grow at weak coupling, but strongly decreases at strong coupling.

We also studied the spatial correlations in these systems at various temperatures and coupling strengths. We then mapped the results of the two-component case to the results of lattice simulations of color magnetic monopoles in pure-gauge $SU(2)$, and find a very good agreement. This comparison allowed us to fix the ``physical line" in the parameter space of our effective model in $SU(2)$ gauge theory, at and above $T_\text{c}$. As a result of simulations, we believe that  a two-component Coulomb quantum Bose gas model accurately replicates monopole behavior  seen on the lattice.

 We have also determined the  monopole contribution  to the overall thermodynamics (energy density) of the thermal matter, at and above $T_\text{c}$ in pure-gauge $SU(2)$, and made estimates for $SU(3)$ and QCD theories. We concluded that the monopoles possibly dominate the thermodynamics just above $T_\text{c}$ in the case of $SU(2)$ and $SU(3)$. We speculate that the same is true in QCD with light quarks, although the questions related with properties of monopole-quark composites have not yet been addressed.

Now having an effective quantum model for magnetic sector of the gauge theories, one may think of its applications beyond quantities calculable in Euclidean-time framework, in particular, to the transport properties of hot hadronic matter. The studies on the impact of monopoles on  QGP viscosity  $\eta$ have been carried out in \cite{Ratti:2008jz}, but those only consider transport cross section of binary collisions, not a full many-body theory. The role of jet-monopoles scattering in another transport parameter -- $\hat q$ -- has been studied phenomenologically in \cite{Xu:2014tda, Xu:2015bbz}.  Both studies suggested that monopoles are the main degree of freedom contributing to $\eta$ and $\hat q$ near and above $T_\text{c}$. Clearly, more quantitative studies of these issues can now be carried out in the framework of our Coulomb Bose gas model.

\vskip 1cm

{\bf Acknowledgements.}
The authors would like to thank the Institute for Advanced Computational Science (IACS) at Stony Brook University for the use of the LI-red computational cluster. AR would like to thank M. Mace and A. Mazeliauskas for helpful discussions, and D. Teaney for making his computer available for running smaller simulations.
This work was supported in part by the U.S. D.O.E. Office of Science,  under Contract No. DE-FG-88ER40388.

\appendix
\section{Path Integrals, Density Matrices, and the Partition Function}\label{app_intro}

In quantum mechanics, the density matrix is related to the path integral by
\eq{
\rho(x_i,x_f,t) = \int \mathcal{D}x(t) \exp\left\{\frac{i}{\hbar}S[x]\right\}\,,
}
where $S[x]$ is the functional action of the particular path $x$. For the purposes of this paper, we will use natural units, where $\hbar$ is set to unity. This path through time corresponds to the usual quantum mechanical evolution operator, $\exp\{i \hat{H} t\}$, where $\hat{H}$ is the Hamiltonian of the system. 

To find the density matrix in finite temperature statistical mechanics, we transform to Euclidean time, $\tau = it$, with periodicity $\beta = T^{-1}$, setting the Boltzmann constant $k_B$ to unity. Then, the thermal density matrix is given by
\eq{
\rho(x_i,x_j,\beta) =  \int \mathcal{D}x(t) \exp\left\{-S_\text{E}[x]\right\}\,,
}
where $S_\text{E}$ is the Euclidean action. The density matrix can also be decomposed in terms of its energy eigenstates
\al{
\rho(x_i,x_j,\beta) = \sum_n \psi_n^*(x_i) \psi_n(x_j) \exp\left\{-E_n \beta \right\} \\ = \braket{\psi_n|\hat{\rho}|\psi_n}\,, \nonumber
}
where $\hat{\rho} = \exp\{-\beta \hat{H}\}$. The density matrix has the property of \textit{squaring},
\eq{
\rho(x_i, x_j,\beta) = \int \dd x_n \rho(x_i,x_n,\beta/2)\rho(x_n,x_j,\beta/2) \,,
}
which then allows the decomposition of the density matrix,
\al{
\rho(x_0,x_M, \beta) &= \int \dd x_1... \dd x_{M-1} \\& \times \rho(x_0,x_1,\tau)...\rho(x_{M-1},x_M,\tau)\nonumber\,,
}
where $\tau = \beta/M$.

If we consider periodic paths, such that $x_i = x_j$, we have that the partition function is
\eq{
Z = \sum_n e^{-\beta E_n} = \Tr[e^{-\beta \hat{H}}] = \Tr[\hat{\rho}] \,,
}
and that the expectation value of the operator $\mathcal{O}$ is
\eq{
\braket{\mathcal{O}} = \frac{\Tr[\mathcal{O}\hat{\rho}]}{\Tr[\hat{\rho}]} = \frac{\Tr[\mathcal{O}\hat{\rho}]}{Z}\,.
}

For a many body system of $N$ particles, we generalize the density matrix as $\rho(R_i,R_j,\beta)$, where $R_i = \{x_1,...,x_N\}$ and  $R_j = \{x^\prime_1,...,x^\prime_N\}$.

In general, the Hamiltonian operator of a system is the sum of the kinetic and potential energy operators, $\hat{H} = \hat{T} + \hat{U}$, and therefore we have that 
\eq{
e^{-\tau \hat{H}} = e^{-\tau(\hat{T}+\hat{V})} = e^{-\tau \hat{T}} e^{-\tau \hat{V}}e^{-\frac{\tau^2}{2} [\hat{T},\hat{V}]} \,.
} 
If $\tau$ is small, then we have the \textit{primitive approximation},
\eq{
 e^{-\tau(\hat{T}+\hat{V})} \approx e^{-\tau \hat{T}} e^{-\tau \hat{V}} \,.
 }
The Trotter formula tells us that this approximation becomes exact in the limit $\tau\rightarrow 0$. Using the fact that $\beta = M \tau$, we have that
\eq{
e^{-\beta(\hat{T}+\hat{V})} =  \lim_{M\to\infty} \left[e^{-\tau \hat{T}} e^{-\tau \hat{V}}\right]^M\,,
}
such that the kinetic and potential actions can be separated and treated individually. Provided $M$, the number of time steps per period of the Matsubara circle, is large enough, we can use this approximation in our numerical simulation to good accuracy. Therefore, in the primitive approximation, the density matrix is given by
\al{
\rho(R_i,R_{i+1},\tau) &= \braket{R_i|e^{-\tau \hat{T}} e^{-\tau \hat{V}}|R_{i+1}} \\&= \braket{R_i|e^{-\tau \hat{T}}|R_{i+1}} e^{-\tau V}\nonumber \,.
}
This quantity describes the degrees of freedom between time slice $i$ and $i+1$. The kinetic matrix element for $N$ particles can be computed using the eigenfunction expansion of the kinetic operator,
\eq{
\braket{R_i|e^{-\tau \hat{T}}|R_{i+1}} = \frac{1}{(4 \pi\lambda\tau)^{3N/2}}\exp{\left\{-\frac{(R_i-R_{i+1})^2}{4\lambda\tau}\right\}}\,,
}
where $\lambda = (2m)^{-1}$.

In the case of $N$ identical bosons, we must also account for permutations of the particles,
\eq{
\rho(R_i,R_j,\beta) = \frac{1}{N!} \sum_P \braket{R_i|e^{-\beta \hat{H}}|P R_j} \,,
}
where $P$ is the permutation operator. The partition function in the primitive approximation is then

\al{
Z = &\frac{1}{N! (4\pi\lambda\tau)^{3N/2}}\nonumber\\&\times\prod_{i = 1}^{M-1} \prod_{n = 1}^N \sum_P \int d x_{i,n}\\&\times\exp{\left\{-\frac{(x_{i,n}-x_{P,i+1,n})^2}{4\lambda\tau} - \tau V(x_{i,n})\right\}} \nonumber \,.
}

\section{Details of the Numerical Simulations}\label{app_details}

The numerical simulations were carried out using a PIMC code written in C++, and the analysis of the output data was done in Python and Mathematica. The algorithm for the PIMC code was the traditional algorithm, with structures and Monte Carlo moves as outlined in \cite{RevModPhys.67.279}. Initially, we implemented the worm algorithm, \cite{PhysRevE.74.036701}, but for large values of the Coulomb coupling, we ran into problems with the acceptances of either the removal or insertion of particles into the system, so we were forced to revert to the traditional algorithm.

\subsection{Metropolis Method}

The path-integral Monte Carlo algorithm is based on Metropolis Monte Carlo (MMC) \cite{Metropolis:1953am}. In MMC, the state of a system is sampled by proposing a random change of state and accepting or rejecting this change based on a probability distribution that is a function of the state. In the case of PIMC, the probability distribution is given in terms of the Euclidean action associated with the state
\eq{
\pi(R) = e^{-S_E(R)}\,.
}
If we propose to move to a new state $R'$, we can compute the acceptance probability of the move
\eq{
A(R\rightarrow R') = \text{min} \left[1, \frac{\pi(R')T(R'\rightarrow R)}{\pi(R)T(R'\rightarrow R)}\right]\,,
}
where $T(x\rightarrow x')$ is the transition probability from state $x$ to state $x'$.

Each particle is represented by a Markov chain, with a \textit{bead}, i.e. physical location, in each of the $M$ time-slices in Matsubara time; the kinetic action is in the ``links" between the beads of the same particle, and the potential action is between the beads in the same time-slice. One then samples changes in locations of these beads, accepting moves using the above probabilities based on the action, and records configurations from which one can sample the partition function and other related thermodynamical quantities.

Further details on the PIMC method, including details on the update moves of the paths, can be found in the all-encompassing review by Ceperley,  \cite{RevModPhys.67.279}.

\subsection{Ewald Summation}

Numerical simulations of Coulomb systems are notoriously difficult to carry out, due to the long range nature of the forces. In computing the potential action for this study, we use the primitive approximation described above and compute the pair-potential for each pair of particle in the simulation. Computing this quantity in position space for periodic boundary conditions is feasible and accurate for short-range potentials, such as that for $^4He$. 

The electrostatic potential for a Coulomb interaction is
\eq{
\Phi_\text{Coulomb}(x_{i,n}) = \sum_\textbf{l} \sum_m \frac{q_m}{|x_{i,n}-x_{i,m}+\textbf{l}|}\,, 
}
where $q_m$ and $x_{i,m}$ are the charge and location of the $m$th particle in the $i$th time-slice, respectively, and \textbf{l} is the vector that corresponds to the periodic image of each particle in space; the $m=n$ case is excluded only for $\textbf{l}=0$. Coulomb interactions, however, such as those of magnetic monopoles, are long-range, and therefore need, in the case of periodic boundary conditions, many images to be accurate. These sums therefore converge slowly and are not good for computational purposes. 

Instead of summing solely in position space, we break up the sum into two rapidly converging pieces, one in position space and the other in reciprocal space, via a technique called Ewald summation \cite{andp.19213690304,TOUKMAJI199673}. Written in this form, and making the conventional choice of the complementary error function for the real sum, the Coulomb potential is
\al{
\Phi_\text{Coulomb}&(x_{i,n}) = \sum_\textbf{l} \sum_m \frac{q_m \text{erfc}(\alpha_\text{cut}|x_{i,n}-x_{i,m}+\textbf{l}|)}{|x_i-x_j+\textbf{l}|} \nn \\
&+ \frac{4\pi}{V}\sum_{\textbf{k}\neq0} \sum_{j} \frac{q_m \exp{\left[\frac{-\textbf{k}^2}{4\alpha_\text{cut}}\right]}}{\textbf{k}^2} \exp{[i \textbf{k} (x_{i,n}-x_{i,m})]} \\
&+ \frac{2\alpha_\text{cut}}{\sqrt{\pi}} q_n \nn \,,
}
where \textbf{k} is the wave vector in reciprocal space and $V$ is the volume of the box. The parameter $\alpha_\text{cut}$ is known as the splitting parameter; it determines the cutoffs for the position and reciprocal space sums. For the simulations with a core, we included a repulsive potential of the form,
\eq{
V_\text{core}(r) = \frac{1}{(5r)^{10}}\,.
}

\subsection{Simulations and Analysis}

Internal energy -- kinetic and potential -- was found using the primitive and virial estimators, both summarized in \cite{RevModPhys.67.279}, and worked out in detail in \cite{Graves:2014}. 

The Coulomb simulations were carried out for 8, 16, 32, and 64 particles, for both the one- and two-component cases. For each simulation, there were 32 imaginary time-slices. For the one-component simulations, we tested temperatures (in our units) from 1.6 to 5.1, in intervals of 0.1, while for the two-component cases, we looked at temperatures in the range of 0.4 to 5.1 in intervals of 0.1, and the range of 6 through 9 in intervals of 1. For the each simulation case -- particle number, temperature, and coupling --  we ran three trials collecting 10000 Monte Carlo configurations post-equilibration, the data from which was then binned and analyzed by a Python script. Error on the raw data was computed using the Jackknife sampling method \cite{Gattringer}. The fits for the data and errors on the fits were computed using Mathematica's \verb|NonlinearModelFit|.

\end{document}